\documentclass[12pt]{article}
\pdfoutput=1
\usepackage{jheppub}
\usepackage{amssymb,amsmath,amstext,amsfonts}
\usepackage{bbold,ulem,bm}
\usepackage{graphics}
\usepackage{mathtools}
\usepackage{hyperref}
\usepackage{color}

\usepackage{graphicx}
\usepackage{subcaption}
\usepackage{mwe}
\usepackage{mathrsfs}
\usepackage{multirow}

\newcommand{\beq}{\begin{equation}}
\newcommand{\eeq}{\end{equation}}
\newcommand{\bea}{\begin{eqnarray}}
\newcommand{\eea}{\end{eqnarray}}
\newcommand{\be}{\begin{equation}}
\newcommand{\ee}{\end{equation}}
\newcommand{\bal}{\begin{aligned}}
\newcommand{\eal}{\end{aligned}}

\def\nn{\nonumber}

\def\M{M_{{\rm Pl}}}
\def\etaperp{\eta_{\perp}}

\def\c{|c_s|}

\def\eq{{\rm eq}}
\def\orth{{\rm orth}}
\def\flat{{\rm flat}}

\def\a{\alpha}

\def\r{\gamma}
\def\t{\theta}
\def\xs{x}
\def\Km{k_{ m}}
\def\kt{\tilde k}

\begin{document}

\title{Flattened non-Gaussianities from the effective field theory of inflation with imaginary speed of sound} 

\author[a]{Sebastian Garcia-Saenz}
\affiliation[a]{Institut d'Astrophysique de Paris, GReCO, UMR 7095 du CNRS et de Sorbonne Universit\'e, 98bis boulevard Arago, 75014 Paris, France}
\author[a]{and S\'ebastien Renaux-Petel}

\emailAdd{sebastian.garcia-saenz@iap.fr}
\emailAdd{renaux@iap.fr}

\abstract{Inflationary perturbations in multi-field theories can exhibit a transient tachyonic instability as a consequence of their non-trivial motion in the internal field space. When an effective single-field description is applicable, the resulting theory is characterized by fluctuations that propagate with an {\it imaginary} speed of sound. We use the effective field theory of fluctuations to study such a set-up in a model-independent manner, highlighting the peculiarities and subtleties that make it different from the standard case. In particular, perturbations feature exponentially growing and decaying modes whose relative amplitude is undetermined within the effective field theory. Nevertheless, we prove that in an interesting limit the dimensionless bispectrum is in fact universal, depending only on the speed of sound and on the cutoff scale that limits the validity of the effective theory. Contrary to the power spectrum, we find that the bispectrum does not display an exponential enhancement. The amplitude of non-Gaussianities in the equilateral configuration is similar to the one of conventional models, but it is enhanced in flattened configurations in a way that is ultraviolet sensitive.}

\maketitle


\section{Introduction} \label{sec:intro}

The predictions of single-field inflation boast a remarkable agreement with the most recent experimental data \cite{Ade:2015lrj,Ade:2015ava}. Yet from a fundamental physics point of view there is no compelling reason to expect that only one field was dynamically active during the inflationary phase, and indeed top-down scenarios of the very early universe typically predict the presence of multiple scalars in addition to the inflaton \cite{Baumann:2014nda,Yamaguchi:2011kg,Linde:2005ht}. It is thus natural to regard the single-field description of inflation as an effective one, arising from the existence of a mass hierarchy between the inflaton and the other fields, which may therefore in practice be integrated out (see \textit{e.g.}~\cite{Tolley:2009fg,Cremonini:2010ua,Achucarro:2010da,Baumann:2011su} for early works). As usual in effective field theory (EFT), the importance of the effects of the heavy fields in the low-energy dynamics is measured by their energy scale. In the context of inflation, this scale need not be extremely large compared to the Hubble scale, and hence it is not unlikely that the heavy modes may lead to sizable effects in the single-field EFT description.

One such effect is that the curvature perturbation will propagate with an effective speed of sound $c_s$ that can differ from the speed of light. The existence of a non-relativistic dispersion relation should of course be expected on general grounds from the spontaneous breaking of time translations by the inflationary background.\footnote{The dispersion relation need not even be linear, see \textit{e.g.}~\cite{ArkaniHamed:2003uz,Baumann:2011su,Ashoorioon:2011eg,Gwyn:2012mw,Gwyn:2014doa,Ashoorioon:2018uey}.} What is interesting is that $c_s$ may be significantly small compared to one, a fact that can lead to important observational signatures, prime among which are large primordial non-Gaussianities with $f_{NL}\sim1/c_s^2$. The bispectrum of cosmological perturbations thus offers a unique window into physics at energies above the Hubble scale (see \textit{e.g.}~\cite{Wands:2010af,Chen:2010xka,Wang:2013eqj,Renaux-Petel:2015bja} for reviews on primordial non-Gaussianities).

An interesting twist to the story is that $c_s$ can in principle be {\it imaginary}. Indeed, in the recently proposed sidetracked inflation scenario \cite{Garcia-Saenz:2018ifx} following the geometrical destabilization of inflation \cite{Renaux-Petel:2015mga,Renaux-Petel:2017dia}, we displayed a concrete realization of an effective single-field theory, arising from a two-field model, that had precisely this property. This is made possible by the fact that the heavy field that is integrated out---the entropic mode---exhibits a transient tachyonic instability that translates into $c_s^2<0$ in the EFT picture. In that case, an imaginary speed of sound is therefore simply a manifestation of this transient amplification of fluctuations, and not a pathology of the underlying UV theory---indeed, there is no fundamental ghost in the two-field models studied in \cite{Garcia-Saenz:2018ifx}. Nevertheless, the resulting tachyonic growth of the fluctuations gives rise to some interesting predictions, which are moreover quite different from those of other inflationary scenarios. First, the curvature power spectrum experiences an exponential growth before the time of sound horizon crossing, leading to a very suppressed tensor-to-scalar ratio. On the other hand, such an exponential growth was shown to be absent in the bispectrum, which was numerically computed in the full two-field description using the transport approach (see \textit{e.g.}~\cite{Dias:2015rca,Dias:2016rjq,Mulryne:2016mzv,Seery:2016lko,Ronayne:2017qzn,Butchers:2018hds} for recent works). However, its amplitude was relatively large, and moreover its shape was found to be markedly different from the usual equilateral one, in particular with a large amplitude in flattened configurations and a large correlation with the orthogonal template.

The goal of the present paper is to derive general results for the size and shape of the bispectrum in imaginary sound speed scenarios. We do so in a model-independent manner by using the EFT of fluctuations \cite{Creminelli:2006xe,Cheung:2007st}, working in the decoupling limit and at lowest order in derivatives. We highlight two features of effective descriptions in terms of an imaginary sound speed. The first is that, since the instability of any underlying UV theory must be transient, such descriptions have to break down at high enough energies. We thus incorporate an appropriate UV cutoff that takes into account the fact that the EFT cannot be infinitely extrapolated towards the past. We also pay particular attention to the quantization of such systems, which plays a central role in our computation. Under the above hypotheses, our results apply to any scenario admitting an effective single-field description for the fluctuations with an imaginary sound speed.\footnote{Although to our knowledge such description has so far only been made explicit in \cite{Garcia-Saenz:2018ifx}, from the results of \cite{Brown:2017osf,Mizuno:2017idt} we conjecture that the same effects are at play in the model of hyperinflation.} The upshot is that such theories lead to simple universal predictions for the bispectrum that can be potentially distinguished from other set-ups. Although we will see that the {\it amplitude} of the non-Gaussianities in the equilateral limit is characterized by $f_{NL}\sim 1/|c_s|^2$, like in conventional frameworks with reduced speed of sound, their {\it shape} is quite distinct, with an enhancement of the bispectrum in flattened configurations, in a way that is sensitive to the UV cutoff of the EFT. This is reminiscent of models with excited initial states (see \textit{e.g.}~\cite{Chen:2006nt,Holman:2007na,Meerburg:2009ys,Meerburg:2009fi,Agarwal:2012mq}), and we explain the similarities and differences between the two frameworks.

In the next section, we briefly review the EFT of fluctuations, and we discuss the quantization and the power spectrum of imaginary sound speed models. We compute the bispectrum and study its amplitude and shape in section \ref{sec:bispectrum}, and discuss our results in section \ref{Discussion}.

\section{Set-up} \label{sec:set-up}

\subsection{Effective field theory of fluctuations}

Our starting point will be the simplest form of the action built from the effective field theory of inflation \cite{Creminelli:2006xe,Cheung:2007st}, or more precisely the EFT of fluctuations generated in single-clock inflation. We refer the reader to the above reference for details, and simply do a brief review of its construction here.

In the unitary gauge where time diffeomorphisms have been fixed so that the clock is unperturbed, there are no matter fluctuations but only metric fluctuations. The most general effective action is then constructed by writing down all operators that are functions of the metric 
fluctuations and invariant under time-dependent spatial diffeomorphisms. It reads, about a spatially flat FLRW background:
\bea
S=\int d^4x \sqrt{-g} \left[ \frac{\M^2}{2} R+\M^2 \dot H g^{00} -\M^2 (3 H^2+{\dot H}) +F(\delta g^{00},\delta K_{\mu \nu}, \delta R_{\mu \nu \rho \sigma};\nabla_{\mu};t) \right] \nn
\eea
where a dot represents the time derivative with respect to the cosmic time $t$, $H \equiv \dot{a}/a$ is the Hubble parameter, $\delta g^{00} \equiv g^{00}+1$, $\delta K_{\mu \nu}$ (respectively $\delta R_{\mu \nu \rho \sigma}$) is the fluctuation of the extrinsic curvature of constant time surfaces (respectively of the 4-dimensional Riemann tensor) and where $F$ starts quadratic in its arguments $\delta g^{00}$, $\delta K_{\mu \nu}$ and $\delta R_{\mu \nu \rho \sigma}$. In the simplest EFT at lowest order in derivatives, one only allows operators involving powers of $\delta g^{00}$, namely, up to cubic order in fluctuations, 
\bea
F&=&\frac{1}{2} M_2(t)^4 (\delta g^{00})^2+\frac{1}{3!} M_3(t)^4 (\delta g^{00})^3\,.
\eea
The Goldstone boson $\pi$ associated with the spontaneous breaking of time-translation invariance can be explicitly reintroduced with the St\"uckelberg trick, restoring full time-diffeomorphism invariance through the replacements $t \to t+\pi$ and $g^{00} \to \partial_{\mu}(t+\pi) \partial_{\nu}(t+\pi) g^{\mu \nu}$. Working in the decoupling limit where $\pi$ decouples from the gravitational sector allows us to neglect the complications of the mixing with gravity,\footnote{Like for $c_s^2>0$, the decoupling limit is not applicable after sound Hubble crossing, but one can use the simple relation $\zeta=-H \pi$ between $\pi$ and the comoving curvature perturbation $\zeta$ to follow the evolution of modes until sound Hubble crossing, after which one makes use of the constancy of $\zeta$.} resulting in the simple transformation law
\be
\delta g^{00} \to -2 {\dot \pi}-{\dot \pi}^2+\frac{(\partial_i \pi)^2}{a^2}\,.
\ee
At leading order in a slow-varying approximation, or equivalently by assuming that $\pi$ enjoys an approximate shift symmetry, one can neglect the time-dependence of all the $M_n(t)$ as well of $H$ and $\dot{H}$. One thus obtains, up to cubic order in $\pi$,
\bea
S_{{\rm DL}}&=& \int {\rm d} t \, {\rm d}^3 x\, a^3 \left[ \M^2 \dot H (\partial_{\mu} \pi)^2+2 M_2^4\left({\dot \pi}^2-{\dot \pi} (\partial_{\mu} \pi)^2\right)
-\frac{4 M_3^4}{3}  {\dot \pi}^3  \right]\,,
\eea
where $(\partial_{\mu} \pi)^2 \equiv -{\dot \pi}^2+(\partial_i \pi)^2/a^2$ is evaluated on the background metric and the comoving curvature perturbation simply reads $\zeta=-H \pi$ at linear order and at leading order in a slow-varying approximation. As a result of the non-linearly realized spontaneously broken symmetry of time diffeomorphism invariance, a non-vanishing $M_2$ introduces both a non-trivial sound speed, such that
\be
\frac{1}{c_s^2}-1 \equiv -\frac{2 M_2^4}{\M^2 {\dot H}}\,,
\ee
and cubic interactions. By defining $A/c_s^2 \equiv -1+\frac23 \left(\frac{M_3}{M_2}\right)^4 $ and $\epsilon \equiv -\dot{H}/H^2>0$, the decoupling Lagrangian is put in the simple form
\bea \label{eq:effective goldstone action}
S_{{\rm DL}}= \int {\rm d} t \, {\rm d}^3 x\, a^3  \M^2  \epsilon H^2 \left[\frac{1}{c_s^2} \left({\dot \pi}^2-c_s^2\frac{(\partial_i \pi)^2}{a^2}\right)-\left(\frac{1}{c_s^2}-1 \right)\left(\frac{{\dot \pi}(\partial_i \pi)^2}{a^2}+\frac{A}{c_s^2} {\dot \pi}^3 \right) \right]\,.
 \label{S-pi}
\eea
where all parameters are taken to be constants, and $a \propto e^{H t}$. We will keep $A$ arbitrary, keeping in mind that $A$ of order one is technically natural as the operators in ${\dot \pi}^3$ and $\dot \pi (\partial_i \pi)^2$ then introduce the same strong coupling scale \cite{Senatore:2009gt}. The power spectrum and the primordial non-Gaussianities originating from the action \eqref{S-pi} are well known when $c_s^2$ is positive \cite{Chen:2006nt,Senatore:2009gt}. Here, on the contrary, we will consider the situation in which $c_s^2$ is negative. We thus write $c_s^2 =- \c^2$, and refer to this situation as a framework with an effective imaginary speed of sound, as it formally corresponds to replacing $c_s$ by $i \c$.

\subsection{Quantization and power spectrum} \label{sec:quantization}

While it may be surprising at first sight to consider the action \eqref{S-pi} in the regime where $c_s^2<0$, as the kinetic energy of $\pi$ is negative then, it can perfectly make sense as a low-energy EFT, as long as we specify its range of applicability. As we explained in the introduction as a motivating example, an effective action with an imaginary speed of sound can indeed be derived from a more fundamental and perfectly healthy theory, like in two-field models after having integrated out a heavy tachyonic field, with a background trajectory deviating strongly from a field space geodesic \cite{Garcia-Saenz:2018ifx}.\footnote{In appendix \ref{sec:appendix} we give an explicit derivation of the single-field EFT arising in the low-energy regime of a generic two-field model of inflation.} In such a system, the EFT becomes valid, for a given scale $k$, when the physical momentum $k/a$ becomes negligible compared to the mass of the heavy tachyonic field. In a similar fashion, we will work under the assumption that the action \eqref{S-pi} with $c_s^2<0$ is valid, for a given scale $k$, when $k \c/(aH)$ drops below a $k$-independent dimensionless quantity, that we call $\xs$, and that is larger than unity. The quantity $\xs$ measures from how deep inside the sound horizon the EFT is applicable, and by introducing the conformal time $\tau$ such that $a \simeq -1/(H \tau)$, our EFT is thus valid, for a given scale $k$, for $k \c \tau +\xs> 0$.\\

Despite the fact that the action \eqref{S-pi} seems to describe an essentially classical instability for $c_s^2<0$, the quantum nature of the system will be central for the understanding of the non-Gaussianities it generates. Therefore, we now discuss its quantization, following a standard procedure, but treating the situations with $c_s^2$ positive or negative in a unified manner for pedagogical reasons. The variable $\zeta$ is promoted to a quantum field, decomposed as 
\beq
\label{Fourier}
\hat \zeta (\tau, \vec x)=\int \frac{{\rm d}^3k}{(2\pi)^3} \left\{{\hat a}_{\vec k} \zeta_{k}(\tau) e^{i \vec k.\vec x}
+ {\hat a}_{\vec k}^\dagger \zeta_{k}^*(\tau) e^{-i \vec k.\vec x} \right\},
\eeq
where the $\hat a$ and $\hat a^\dagger$ are annihilation and creation operators, which satisfy the 
usual commutation rules 
\beq
\label{a}
\left[ {\hat a}_{\vec k}, {\hat a^\dagger}_{\vec k'}\right]=(2\pi)^{3} \delta(\vec k-\vec k')\, ,
\quad
\left[ {\hat a}_{\vec k}, {\hat a}_{\vec k'}\right]= 
\left[ {\hat a^\dagger}_{\vec k}, {\hat a^\dagger}_{\vec k'}\right]= 0\, ,
\eeq
and the complex mode function $ \zeta_{k}(\tau)$ verifies the equation of motion
\be
\zeta_k^{''}-\frac{2}{\tau} \zeta_k^{'}+c_s^2 k^2 \zeta_k=0\,.
\ee
Its general solution reads
\beq
\zeta_k(\tau)=\frac{A_k}{k^{3/2}} e^{- i k c_s \tau}(-i k c_s  \tau-1)+\frac{B_k}{k^{3/2}} e^{i k c_s \tau}(i k c_s \tau-1)\,,
\label{general-solution}
\eeq
where $A_k$ and $B_k$ are arbitrary complex constants, and $c_s$ denotes the positive square root of $c_s^2$ when the latter is positive, and $i |c_s|$ when it is negative. \\

In conformal time, the conjugate momentum of $\zeta$ is $p_\zeta=2 a^2 \epsilon \M^2/c_s^2 \zeta^{'}$, where $'={\rm d}/{\rm d} \tau$. The quantization condition $\left[ \hat \zeta (\tau, \vec x), \hat p_\zeta^\dagger (\tau, \vec x)\right]=i$ thus imposes that $ \zeta_{k}(\tau)$ verify
\beq
\zeta_k \zeta^{'*}_k-\zeta'_k \zeta^*_k=\frac{i c_s^2}{2 \epsilon a^2 \M^2}
\eeq
at all times, which leads to the constraint
\be
\hspace{-0.2cm} {\rm Re}[c_s] \left( |A_k|^2 e^{2 k \tau {\rm Im}[c_s]}- |B_k|^2 e^{-2 k \tau {\rm Im}[c_s]}  \right) +2\, {\rm Im}[c_s] {\rm Im}[A_k^* B_k e^{2i k \tau {\rm Re}[c_s]}] =\frac{H^2}{4 \epsilon \M^2}\,.
\label{quantization-universal}
\ee
The consequences of the quantization condition are therefore very different for the two cases of positive and negative $c_s^2$:\\

\textbullet \,\, For $c_s^2$ positive, $ {\rm Im}[c_s]=0$, so that the second term is zero, and one finds the familiar constraint
\be
|A_k|^2-|B_k|^2 =\frac{H^2}{4 \epsilon c_s \M^2}\,.
\label{quantization-standard}
\ee
One is then free to choose $B_k=0$, the Bunch--Davies vacuum, selecting only the positive frequency mode in \eqref{general-solution}, whose amplitude is completely fixed, finding then for the late power spectrum ${\cal P}_{\zeta_k} \equiv k^3/(2 \pi^2) |\zeta_k|^2= H^2/(8 \pi^2 \epsilon c_s \M^2)$.\\

\textbullet \,\,For $c_s^2$ negative, $ {\rm Re}[c_s]=0$, so that the first term in \eqref{quantization-universal} is zero, and one finds
\beq
  {\rm Im}[A_k^* B_k]=\frac{H^2}{8 \epsilon |c_s| \M^2}\,.
\label{quantization-A-B}
\eeq
Here, the quantization condition thus imposes that both $A_k$ and $B_k$ be non-zero. In the following, as the global phase of the mode function is irrelevant, one chooses $A_k$ to be real, so that a more precise statement is that the imaginary part of $B_k$ should be non-vanishing. We also write $A_k=\alpha_k e^{\xs}$ and $B_k=\alpha_k e^{\r_k} e^{i \t_k} e^{-\xs}$, where all parameters are real, so that the mode function reads
\beq
\zeta_k(\tau)=\frac{\a_k}{k^{3/2}} \left( e^{k |c_s| \tau+\xs}(k |c_s| \tau-1)-e^{\r_k} e^{i \t_k} e^{-(k |c_s| \tau+\xs)}(k |c_s| \tau+1) \right)\,.
\label{mode-function}
\eeq
The positive and negative frequency modes of the situation with $c_s^2>0$ are now turned into exponentially growing and decreasing modes (the first and second terms respectively). The physical motivation behind this parameterization is the following. As the EFT starts to be valid at $k |c_s \tau |=\xs$ for the scale $k$, one cannot predict the amplitude of the growing and decaying modes without further input from a UV completion. However, on physical grounds, one can expect them to be excited with similar amplitudes. Hence, we factored out the $e^{\pm \xs}$ terms, so that $e^{\r_k}$, which sets the initial ratio between the amplitudes of the decaying and growing modes, is considered in the following to be an order one number. Eventually, note that with these parameters, the quantization condition \eqref{quantization-A-B} reads
\beq
 \a_k^2 e^{\r_k} \sin(\t_k)=\frac{H^2}{8 \epsilon |c_s| \M^2} \,.
\label{quantization-apha}
\eeq
Taking the limit $k |c_s| \tau \to 0$, one finds from \eqref{mode-function} the final value of the power spectrum:
\beq
{\cal P}_{\zeta_k}= \frac{ \a_k^2}{2 \pi^2} \left[ e^{2\xs} + e^{2 \r_k} e^{-2 \xs} +2e^{\r_k} \cos(\t_k)  \right]\,.
\label{Pzeta-exact}
\eeq
Let us recall that $e^{\r_k}={\cal O}(1)$ and that $\xs$ is positive, so that a soon as $\xs \gtrsim 5$, one is completely dominated by the first term, coming from the exponentially growing mode, so that
\beq
{\cal P}_{\zeta_k} \simeq \frac{ \a_k^2}{2 \pi^2}  e^{2\xs}\,.
\label{Pzeta}
\eeq
In the following, in agreement with the approximate shift symmetry that we used, we will work under the simplifying assumption that the curvature power spectrum \eqref{Pzeta} is scale invariant, \textit{i.e.}~that $\alpha_k$ does not depend on the scale $k$, and we write $\alpha_k\equiv\alpha$ in the following. From the quantization condition \eqref{quantization-apha}, one deduces that $e^{\r_k} \sin(\t_k)$ is also scale-independent, and for simplicity we will assume that the amplitude and the phase are separately scale-independent, hence omitting the explicit subscripts $k$ from now on. These approximations are mainly used to simplify the results, but play no fundamental role in our computation. Notice finally that in the particular set-up of sidetracked inflation \cite{Garcia-Saenz:2018ifx}, we used a heuristic description of the transition from the UV regime to the one of the EFT that was sufficient for our purpose there to model the power spectrum, but that did not take into account the quantization condition \eqref{quantization-apha}. Here, on the contrary, we remain agnostic about the precise amplitude of $\alpha^2$, and hence of the curvature power spectrum. Nevertheless, we will show that under mild assumptions, one can derive universal results about the amplitude and shape of the primordial non-Gaussianities.

\section{Bispectrum from an imaginary speed of sound} \label{sec:bispectrum}

In this section we compute the three-point correlation function of the curvature perturbation: 
\beq
\langle \zeta_{\boldsymbol{k}_1} \zeta_{\boldsymbol{k}_2} \zeta_{\boldsymbol{k}_3} \rangle \equiv (2\pi)^3 \delta(\sum_i \boldsymbol{k}_i) B_{\zeta}(k_1,k_2,k_3)\,, 
\label{Bispectrum}
\eeq
where the bispectrum $B_\zeta$ is a function of the three wavenumbers $k_i=|\boldsymbol{k}_i|$. In particular, we will study the amplitude and the momentum-dependence of the shape function $S$ such that \cite{Babich:2004gb}
 \beq
B_{\zeta} \equiv (2 \pi)^4 \frac{S(k_1,k_2,k_3)}{(k_1 k_2 k_3)^2} A_s^2\,,
\label{shape-def}
\eeq
where $A_s \simeq 2.4 \times 10^{-9}$ denotes the amplitude of the dimensionless curvature power spectrum ${\cal P}_\zeta$ at the pivot scale $k=0.05\,{\rm Mpc}^{-1}$, which reads, according to Eq.~\eqref{Pzeta},
\be
A_s=\frac{\a^2}{2 \pi^2}\,  e^{2\xs}\,.
\label{As}
\ee
Using the in-in formalism (see \textit{e.g.}~\cite{Maldacena:2002vr,Weinberg:2005vy}), the tree-level bispectrum can be computed as
\be
 \langle \zeta_{\boldsymbol{k}_1}(t) \zeta_{\boldsymbol{k}_2}(t) \zeta_{\boldsymbol{k}_3}(t) \rangle=2 \,{\rm Im} \left[ \int_{-\infty(1-i \epsilon)}^t {\rm d} t' \langle 0| \zeta_{\boldsymbol{k}_1}(t) \zeta_{\boldsymbol{k}_2}(t) \zeta_{\boldsymbol{k}_3}(t) H_{(3)}(t') | 0 \rangle \right]\,,
 \label{in-in}
\ee
where the interaction picture cubic Hamiltonian $H_{(3)}$ simply reads $-\int {\rm d}^3x {\cal L}_{(3)}$ here, and for simplicity, we omit indices to denoting that all fields are in the interaction picture. As we already emphasized, an important feature of our calculation is that our EFT is valid only for $k |c_s  \tau | \geq \xs$ for a given scale $k$. Hence, in Eq.~\eqref{in-in}, we can only compute the contribution to the bispectrum starting from the time such that all modes have reached the regime of validity of the EFT, namely from $\tau_i=-\xs/(|c_s|\Km)$ with $\Km \equiv\,{\rm max}(k_1,k_2,k_3)$. We disregard any previous contribution to the bispectrum, which would depend on the model-dependent UV completion of the EFT. In standard set-ups, and when the three modes have comparable orders of magnitude, we expect this contribution, approximately coming from the period when modes are oscillating, to be anyway dwarfed in magnitude by the subsequent contribution that we compute, where all modes experience an exponential growth. On the other hand, in the squeezed limit, one can make use of the single-clock consistency relation \cite{Maldacena:2002vr,Creminelli:2004yq} that predicts a vanishing bispectrum in our approximation of a scale-invariant power spectrum. As we will see, given that our computations result in a vanishing three-point function in the squeezed limit, our shape can thus be considered to be reliable for all type of triangle configurations.

\subsection{Computation of the bispectrum}
\label{computation}

In this subsection we give our result for the bispectrum, treating subsequently each contribution from the two vertices of the cubic action. We begin with the algebraically simpler case of the vertex $\dot{\zeta}^3$, for which we explain in some detail the structure and the subtleties of the computation.\\

{\bf Interaction in $\dot{\zeta}^3$.---} Using Eqs.~\eqref{S-pi}-\eqref{in-in}, with $\zeta=-H \pi$ and $a \simeq -1/(H \tau)$, one readily obtains the contribution to the final bispectrum from the vertex $\dot{\zeta}^3$ as:
\beq
B_{\zeta} \supset \left(\frac{1}{|c_s|^2}+1 \right) \frac{12 A\, \epsilon \M^2}{H^2 |c_s|^2}  \int_{-\xs/(|c_s|\Km)}^{0} \frac{{\rm d} \tau }{\tau} \,{\rm Im}\left[  \zeta_{k_1}(0)  \zeta_{k_2}(0)  \zeta_{k_3}(0) \zeta^{*'}_{k_1}(\tau) \zeta^{*'}_{k_2}(\tau) \zeta^{*'}_{k_3}(\tau) \right]\,,
\label{B1}
\eeq
where 
\beq
\zeta^{'}_k(\tau)=\frac{\a_k}{k^{3/2}} |c_s|^2 k^2 \tau \left( e^{k |c_s| \tau+\xs}+e^{\r_k} e^{i \t_k} e^{-(k |c_s| \tau+\xs)} \right)\,.
\label{mode-function-derivative}
\eeq
There is no difficulty in performing the relevant integrals exactly, but it is cumbersome to write the full result, and it is physically more instructive to discuss the structure of the computation to identify the leading-order result. To this end, note that the growing mode in \eqref{mode-function} and \eqref{mode-function-derivative} comes with a large factor $e^{\xs}$, while the decaying mode comes with an $e^{-\xs}$, hence it is easy to organize the computation by formally counting the powers of $e^{\xs}$. 

Taking into account only the growing mode, the dominant term inside the brackets in \eqref{B1} scales as $e^{6 \xs}$. However, it is real and hence does not contribute to the bispectrum. This explains why in section \ref{sec:quantization}, we paid special attention to the quantization, and in particular to the imaginary part of the decaying mode. The first non-zero contribution to the bispectrum thus comes from inserting one decaying mode in the product of the six mode functions, and thus scales like $e^{4 \xs}$, while other contributions are suppressed by additional powers of $e^{-\xs}$. Hence, in the limit of a large $\xs$ (a statement that will be made quantitative below), we find a dominant contribution proportional to $e^{4 \xs} \alpha^6 e^{\r} \sin(\theta)$. As the dimensionless shape function \eqref{shape-def} involves the ratio between the bispectrum and $A_s^2$, with the latter scaling like $\a^4 e^{4 \xs}$, one finds a result for $S$ proportional to $\a^2 e^{\r} \sin(\theta)$, whose amplitude is fixed by virtue of the quantization condition \eqref{quantization-apha}. Contrary to the power spectrum, and to what a naive power counting might have led to, the bispectrum is therefore not enhanced by $e^{2 \xs}$, but its overall amplitude is simply set by $1/c_s^2-1$, like in conventional models with positive speed of sound squared (we will see though that the result is enhanced in flattened configurations, but only by $\xs^3$ instead of exponentially). Performing the integrals explicitly, one finds the following leading-order result:
\beq\bal
S_{\dot{\zeta}^3}=   \frac{3A}{4} \left(\frac{1}{|c_s|^2}+1 \right)\bigg\{&-\frac{k_1k_2k_3}{(k_1+k_2+k_3)^3}  \\
&+\frac{k_1k_2k_3}{\kt_1^3}\bigg[1-e^{-\xs \kt_1/\Km} \bigg(1+\xs \frac{\kt_1}{\Km}+\frac{\xs^2}{2} \frac{\kt_1^2}{\Km^2}\bigg)\bigg]\bigg\}+(\mbox{2 perm.})
\label{Ssimple}
\eal\eeq
where
\beq
\kt_1\equiv k_2+k_3-k_1\,,
\eeq
and similarly for $\kt_2$ and $\kt_3$. Note that $\kt_i\geq0$ as a consequence of the triangle inequality. \\

It is easy to understand physically the momentum dependence of the various contributions. The first term comes from inserting one decaying mode in the external legs in Eq.~\eqref{B1}. The integrand is then computed by taking the product of three growing modes for the internal legs, hence is proportional to $\tau^2\,e^{(k_1+k_2+k_3)|c_s| \tau}$ and leads to a standard equilateral type result.\footnote{Note that we neglected terms in $e^{-\xs(k_1+k_2+k_3)/\Km}$, which are suppressed at least by $\xs^2 e^{-2 \xs}$ in all triangle configurations.} The fact that this contribution is the same as for $c_s^2$ positive should not come as a surprise as, in that case, the requirement to project into the interacting vacuum state (the $i\epsilon$ prescription in Eq.~\eqref{in-in}) effectively corresponds to performing the integral with $c_s$ turned into $i |c_s|$. The second contribution comes from inserting one decaying mode in the internal legs, changing one $k_i$ into $-k_i$ in the integrand, which becomes proportional to $\tau^2\,e^{(-k_1+k_2+k_3)|c_s| \tau}$ (and permutations), and hence to a shape that is enhance in flattened configurations such that $k_2+k_3=k_1$ (and permutations). 

The result that we obtain is therefore similar in spirit to the bispectrum generated by a small non-Bunch--Davies component in conventional models with positive $c_s^2$ (see \textit{e.g.}~\cite{Chen:2006nt,Holman:2007na,Meerburg:2009ys,Meerburg:2009fi,Agarwal:2012mq}). Note however that in our case, the result \eqref{Ssimple} is not a correction that comes in addition to a standard equilateral type bispectrum, but constitutes the dominant bispectrum itself. In addition, as a direct consequence of the presence of an imaginary speed of sound, the bispectrum does not feature an oscillating behaviour, but rather acquires an exponential dependence in $\xs \kt_1/\Km$ (and permutations). The terms in $e^{-\xs \kt_1/\Km}$ are negligible near the equilateral limit, but they become increasingly important as one approaches the flattened configurations, and are in fact crucial to regularize the apparent divergence in $1/\kt_1^3$ in this limit. Note also that the equilateral type contribution, coming from the correction to the external legs, is numerically smaller than the other contribution, even in the equilateral configuration, and could be neglected for a simpler but qualitatively correct result. 

A word is also useful about the terms that we have neglected in \eqref{Ssimple}. From the structure of the computation, one can realize that they are negligible for all type of triangle configurations, and that they are parametrically suppressed (at least) by $\xs^2 e^{-\xs}$ compared to the leading-order result in Eq.~\eqref{Ssimple}. Let us quote for instance the next-to-leading-order correction in $e^{-\xs}$, in the equilateral configuration for simplicity:
\bea
S_{\dot{\zeta}^3 \,{\rm NLO}}(k,k,k)=\frac{9 A}{4} \left(\frac{1}{|c_s|^2}+1 \right)e^{-\xs} \bigg[&&\hspace{-0.3cm}-(1 + \xs + \xs^2/2) - e^{\r} (2 - 2 \xs + \xs^2) \cos(\theta)  \nn \\ 
   &&\hspace{-0.5cm} +\frac{1}{156}e^{2 \r}(2- 6\xs +   9 \xs^2)(1+2\cos(2 \theta))  \bigg]\,.
\eea
Like the correction to the leading-order power spectrum in Eq.~\eqref{Pzeta-exact}, the corrections depend on $e^{\r}$ and $\cos(\theta)$, but the dominant suppressing factor is not $e^{-2 \xs}$ but rather $\xs^2 e^{-\xs}$. Thus, the quantitative criterion that enables us to derive model independent results for the bispectrum from the EFT is that $\xs^2 e^{-\xs} \ll 1$, which is verified for $\xs \gtrsim 8$. This is the regime that we consider in the remainder of this paper. \\

{\bf Interaction in $\dot{\zeta}(\partial\zeta)^2$.---} Using Eqs.~\eqref{S-pi}-\eqref{in-in}, the contribution to the final bispectrum from the vertex $\dot{\zeta}(\partial\zeta)^2$ reads
\bea
B_{\zeta} &\supset & \left(\frac{1}{|c_s|^2}+1 \right) \frac{4  \epsilon \M^2}{3 H^2} \, \boldsymbol{k}_2 \cdot \boldsymbol{k}_3  \times \nn \\
&&\int_{-\xs/(|c_s|\Km)}^{0} \frac{{\rm d} \tau }{\tau} \,{\rm Im}\left[  \zeta_{k_1}(0)  \zeta_{k_2}(0)  \zeta_{k_3}(0) \zeta^{*'}_{k_1}(\tau) \zeta^{*}_{k_2}(\tau) \zeta^{*}_{k_3}(\tau) \right]+(\mbox{2 perm.})\,.
\label{B2}
\eea
The structure of the calculation is completely analogous to the previous one, only with an algebraically more complicated momentum-dependence. Hence we simply quote the final result, again keeping only leading-order terms:
\beq\bal
\label{Suni}
S_{\dot{\zeta}(\partial\zeta)^2}&=\frac{1}{16} \left(\frac{1}{|c_s|^2}+1 \right)\bigg\{-\frac{k_1k_2k_3}{(k_1+k_2+k_3)^3}\,p_0(k_1,k_2,k_3)+\frac{k_1k_2k_3}{\kt_1^3} \times  \\
&\hspace{-1cm}\bigg[p_0(-k_1,k_2,k_3)-e^{-\xs \kt_1/\Km} \bigg(p_0(-k_1,k_2,k_3)+\xs \frac{\kt_1}{\Km}p_1(-k_1,k_2,k_3)+\frac{\xs^2}{2} \frac{\kt_1^2}{\Km^2} p_2(-k_1,k_2,k_3)\bigg)\bigg] \bigg\}\\
&+(\mbox{2 perm.}) 
\eal\eeq
with
\bea\bal
p_0(k_1,k_2,k_3)&=-12-9\bigg(\frac{k_1}{k_2}+\mbox{5 perm.}\bigg)-\bigg(\frac{k_1^2}{k_2^2}+\mbox{5 perm.}\bigg)+6\bigg(\frac{k_1^2}{k_2k_3}+\mbox{2 perm.}\bigg)\\
&\quad-6\bigg(\frac{k_1k_2}{k_3^2}+\mbox{2 perm.}\bigg)+3\bigg(\frac{k_1^3}{k_2k_3^2}+\mbox{5 perm.}\bigg)+\bigg(\frac{k_1^4}{k_2^2k_3^2}+\mbox{2 perm.}\bigg)\\
p_1(k_1,k_2,k_3)&=-6-5\bigg(\frac{k_1}{k_2}+\mbox{5 perm.}\bigg)+4\bigg(\frac{k_1^2}{k_2k_3}+\mbox{2 perm.}\bigg)\\
&\quad-2\bigg(\frac{k_1k_2}{k_3^2}+\mbox{2 perm.}\bigg)+\bigg(\frac{k_1^3}{k_2k_3^2}+\mbox{5 perm.}\bigg)\\
p_2(k_1,k_2,k_3)&=-2\bigg(\frac{k_1}{k_2}+\mbox{5 perm.}\bigg)+2\bigg(\frac{k_1^2}{k_2k_3}+\mbox{2 perm.}\bigg)\,.
\eal
\eea
The physical origin of the various momentum-dependences is similar to the above case: the first contribution simply comes from inserting a decaying mode in one of the three external legs in \eqref{B2}, and for the same reason as for the other interaction, it has the same equilateral-type shape as the same vertex in models with $c_s^2>0$. The other contribution stems from inserting one decaying mode in the internal legs, turning one $k_i$ into $-k_i$, thus yielding a shape that is enhanced in flattened configurations, where the exponential terms are important to regularize the apparent divergence in $1/\kt_1^3$. Finally, note that contrary to the other interaction \eqref{Ssimple}, the equilateral type shape does not give a contribution in the equilateral configuration that is numerically negligible compared to the one coming from the flattened shape.

\subsection{Shapes and amplitudes}

\begin{figure*}
        \centering
        \begin{subfigure}[b]{0.4\linewidth}
            \centering
            \includegraphics[width=\textwidth]{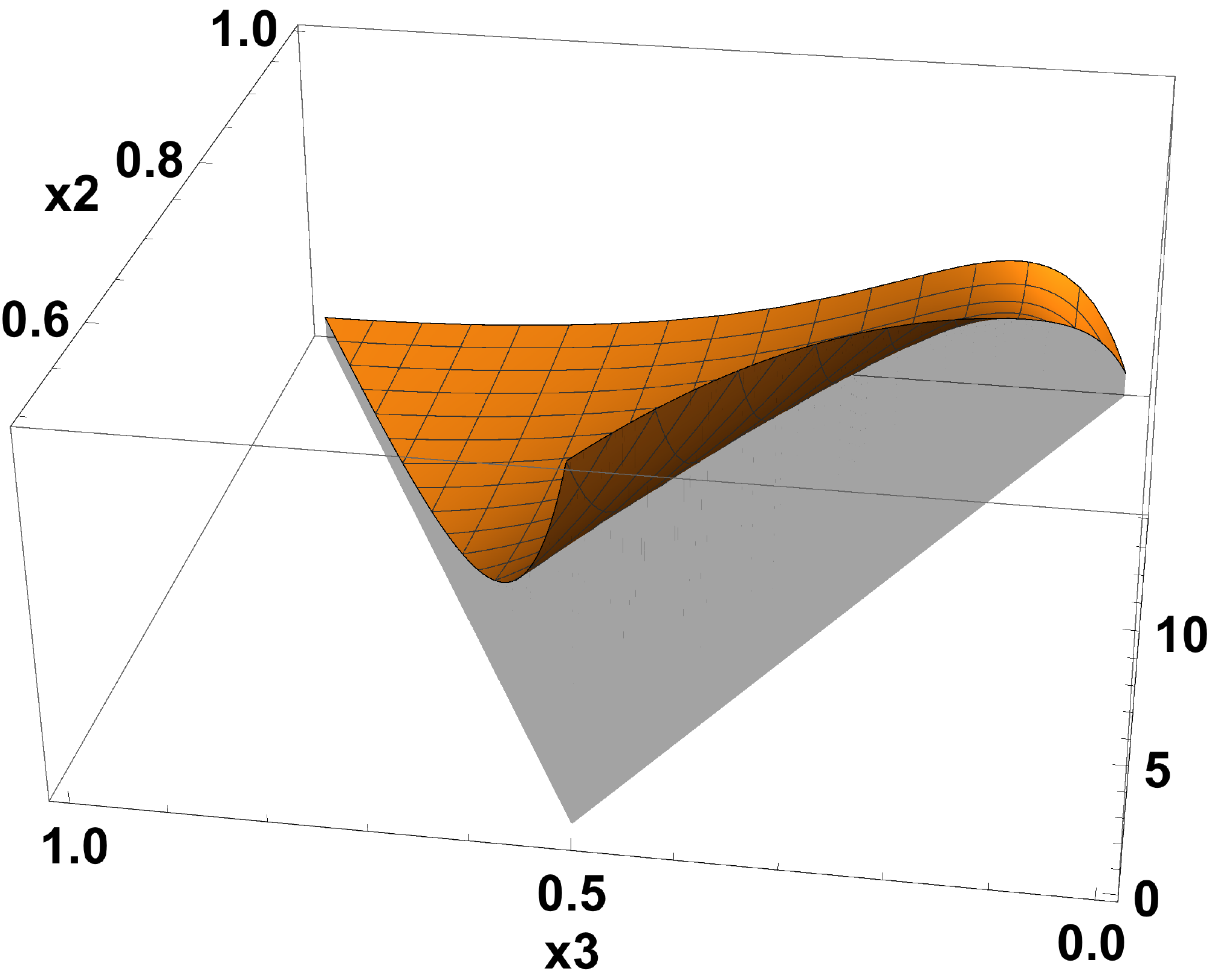}
            \caption{$S_{\dot{\zeta}^3}$ for $\xs=10$.}    
            \label{fig:Ssimple}
        \end{subfigure}
        \hfill
        \begin{subfigure}[b]{0.43\linewidth}  
            \centering 
            \includegraphics[width=\textwidth]{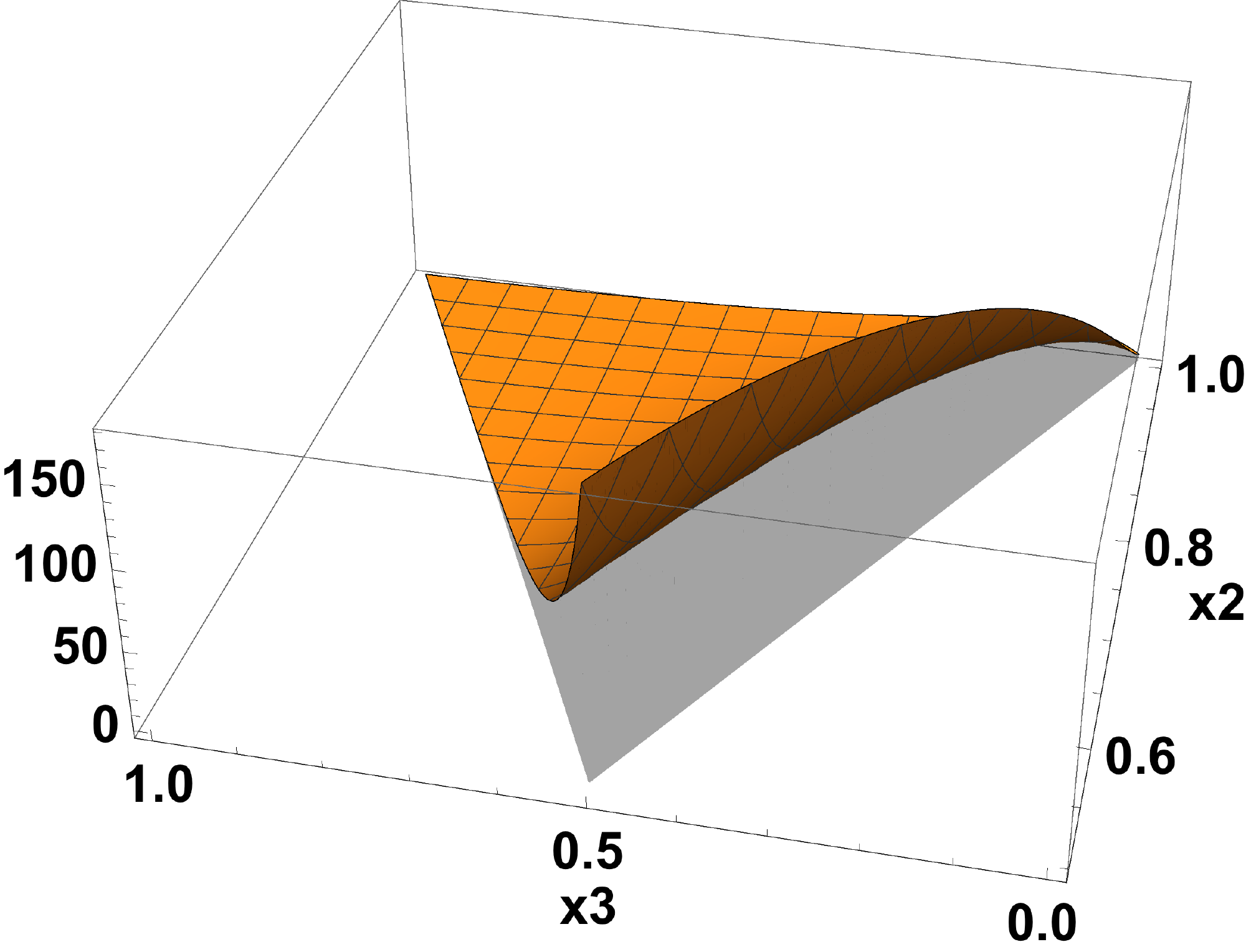}
            \caption{$S_{\dot{\zeta}(\partial\zeta)^2}$  for $\xs=10$.}    
            \label{fig:Suni}
        \end{subfigure}
        \caption{Shapes $S(1,x_2,x_3)$ as a function of $(x_2,x_3)$. We set them to zero outside the region $1-x_2 \leq x_3 \leq x_2$. The shape $S_{\dot{\zeta}^3}$ (respectively $S_{\dot{\zeta}(\partial\zeta)^2}$) is normalized to $1$ (respectively $-1$) in the equilateral configuration $x_2=x_3=1$.}
        \label{fig:Shapes}
    \end{figure*}

In this subsection we study in more detail the amplitudes and the momentum dependences of the two shapes in Eqs.~\eqref{Ssimple}-\eqref{Suni}. For this, we order the $k_i$'s such that $k_3 \leq k_2 \leq k_1$, and we represent the shape information by plotting the two-dimensional functions $S(1, x_2, x_3)$, where $0  \leq x_3 \leq x_2 \leq 1$ and $1 \leq x_2+x_3$ (to satisfy the triangle inequality). The two shapes are represented in Fig.~\ref{fig:Shapes} for the representative value $\xs=10$. Note that $S_{\dot{\zeta}^3}$ (respectively $S_{\dot{\zeta}(\partial\zeta)^2}$) is normalized to $1$ (respectively $-1$) in the equilateral configuration $x_2=x_3=1$. 

As we understood in the last section, the striking feature shared by the two shapes is the large enhancement in flattened configurations $x_2+x_3=1$ compared to the equilateral one (except in the squeezed limit where all shapes vanish).
Indeed, one finds
\bea
S(1,1,1)=  \left(\frac{1}{|c_s|^2}+1 \right) \left( \frac{13A}{6}-\frac{5}{24}\right) 
\eea
for the total shape $S=S_{\dot{\zeta}^3}+A S_{\dot{\zeta}(\partial\zeta)^2}$ in the equilateral limit, while the result in the squashed configuration $(x_2=x_3=1/2)$ reads
\bea
S\left(1,\frac12,\frac12\right)= \frac{1}{128} \left(\frac{1}{|c_s|^2}+1 \right)\bigg[39(A-1)+ 12 \xs^2 + 4 \xs^3(A+1)\bigg]\,,
\label{S-squashed}
\eea
where we kept the dominant terms in each configuration, and it is clear what the contribution from each operator is. One can easily derive an expression for $S(1,x_2,1-x_2)$ in more general flattened triangles, but it is not particularly illuminating, and for simplicity we concentrate on the squashed configuration where each shape is the largest. Let us remark that in Fig.~\ref{fig:Shapes}, if the enhancement in the squashed configuration is more important for $S_{\dot{\zeta}(\partial\zeta)^2}$ than for $S_{\dot{\zeta}^3}$, it is simply because each shape is normalized to $\pm1$ in the equilateral configuration. As one can see from Eq.~\eqref{S-squashed}, for $A={\cal O}(1)$, the two shapes actually contribute equally in the squashed limit, with a dominant term proportional to $\xs^3$ for large $\xs$. The result \eqref{S-squashed}, and the non-trivial dependence on $\xs$ in particular, can be derived by carefully taking the squashed limit from the general expressions \eqref{Ssimple}-\eqref{Suni}, or by considering a squashed configuration from the start in the computations \eqref{B1}-\eqref{B2} of the bispectrum. The argument of the relevant exponential factors being zero in that case, one can see that the $\xs$-enhanced term is proportional to $\int_{-\xs/(|c_s|\Km)}^{0} \tau^2 {\rm d} \tau \propto \xs^3$ for the operator $\dot{\zeta}^3$, while there is also a contribution in $\int_{-\xs/(|c_s|\Km)}^{0} \tau {\rm d} \tau \propto \xs^2$ for the operator $\dot{\zeta}(\partial\zeta)^2$. One can also notice that $S_{\dot{\zeta}^3}$ has the same sign in the equilateral and in the squashed configuration, while$S_{\dot{\zeta}(\partial\zeta)^2}$ changes from negative in the equilateral limit to positive in the squashed one for the relevant values of $\xs$.

\begin{figure*}
        \centering
        \begin{subfigure}[b]{0.4\linewidth}
            \centering
            \includegraphics[width=\textwidth]{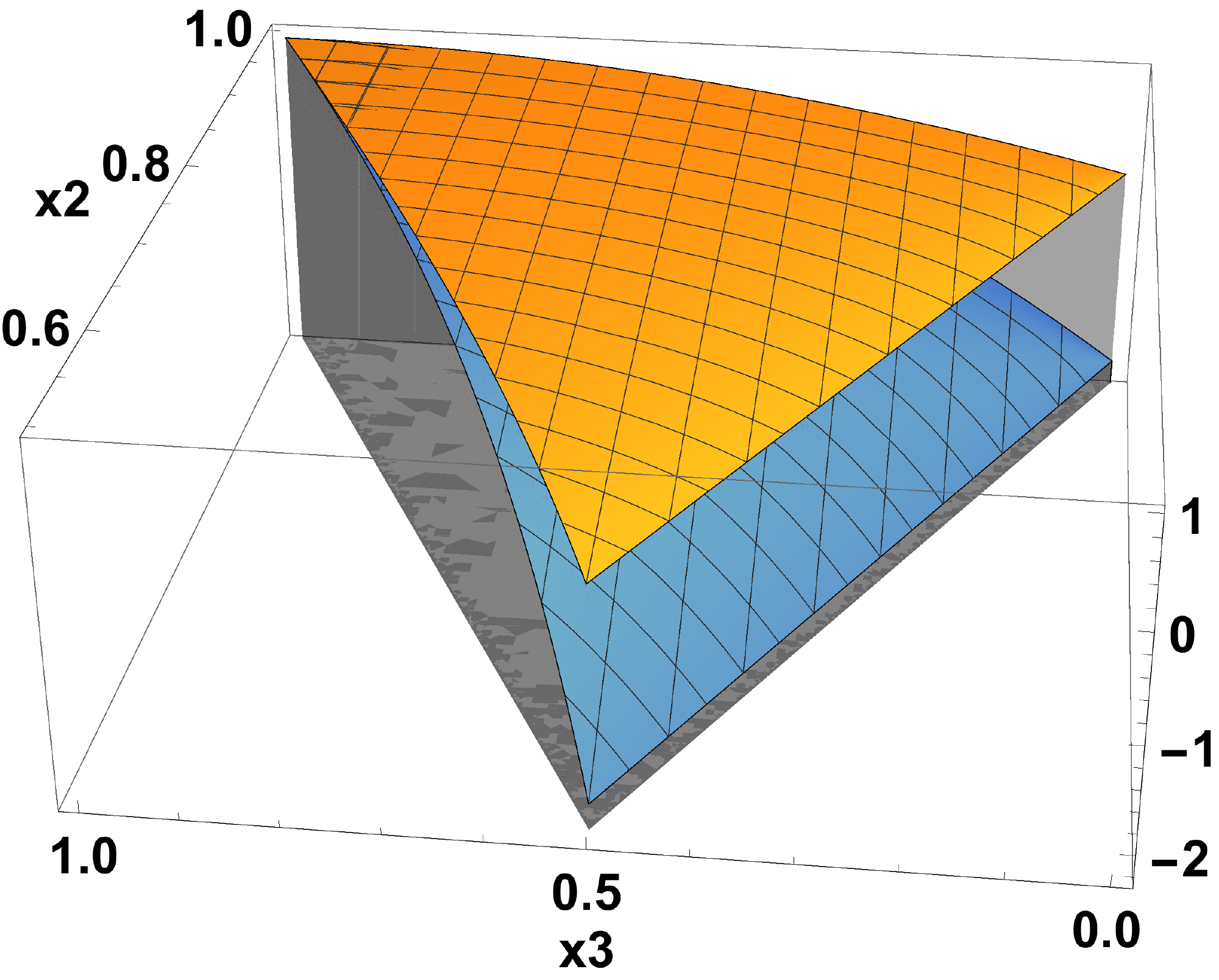}
            \caption{$S^\eq$ (in orange) and $S^\orth$ (in blue), defined in Eq.~\eqref{templates}.}    
            \label{fig:Seq-orth}
        \end{subfigure}
        \hfill
        \begin{subfigure}[b]{0.43\linewidth}  
            \centering 
            \includegraphics[width=\textwidth]{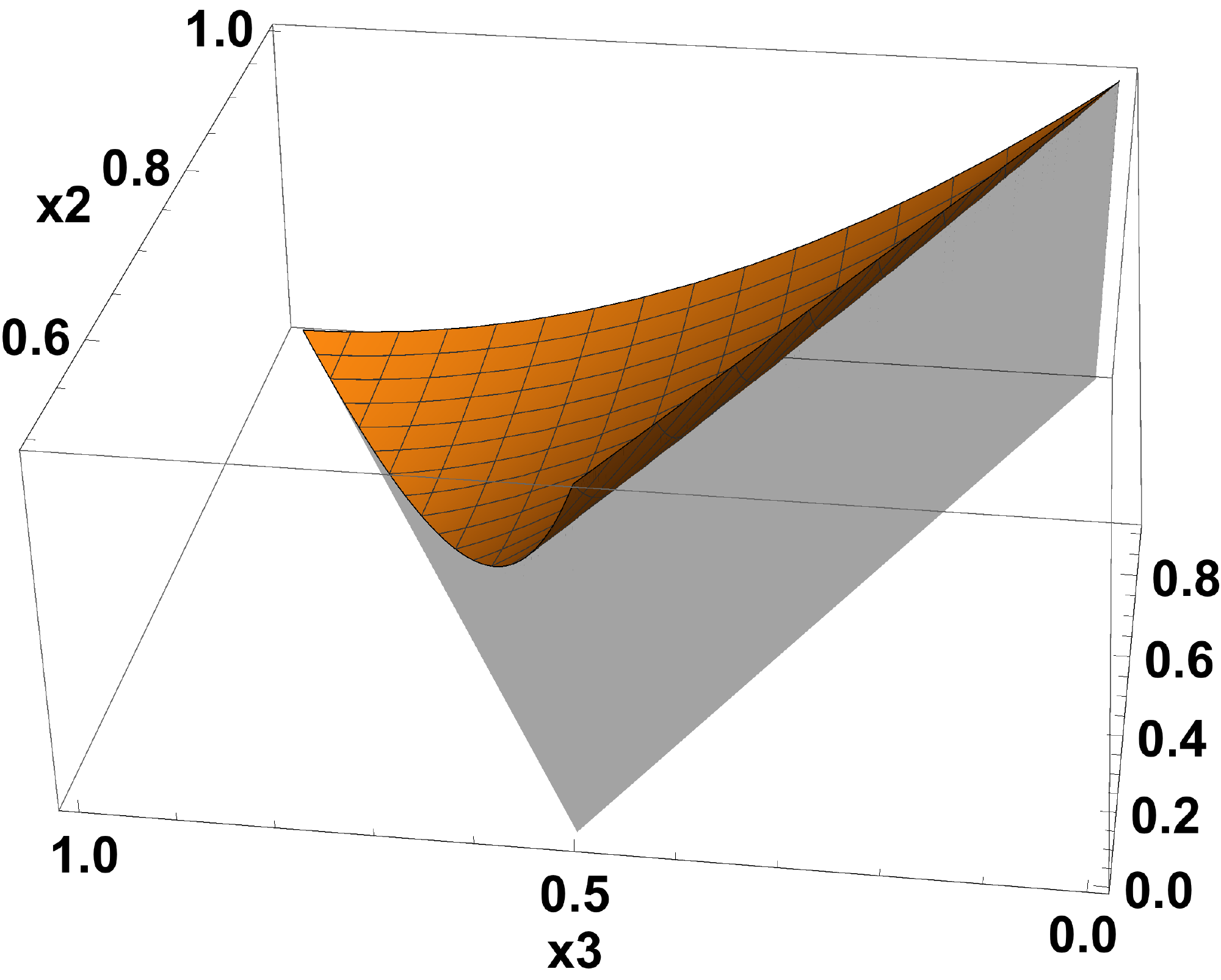}
            \caption{$S^\flat$, defined in Eq.~\eqref{templates}.}    
            \label{fig:Sflat}
        \end{subfigure}
        \caption{Shapes $S(1,x_2,x_3)$ as a function of $(x_2,x_3)$. We set them to zero outside the region $1-x_2 \leq x_3 \leq x_2$, and normalize them to $1$ in the equilateral configuration, except for the flat shape, which vanishes in this limit.}
        \label{fig:templates}
    \end{figure*}

\begin{figure*}
        \centering
        \begin{subfigure}[b]{0.43\linewidth}
            \centering
            \includegraphics[width=\textwidth]{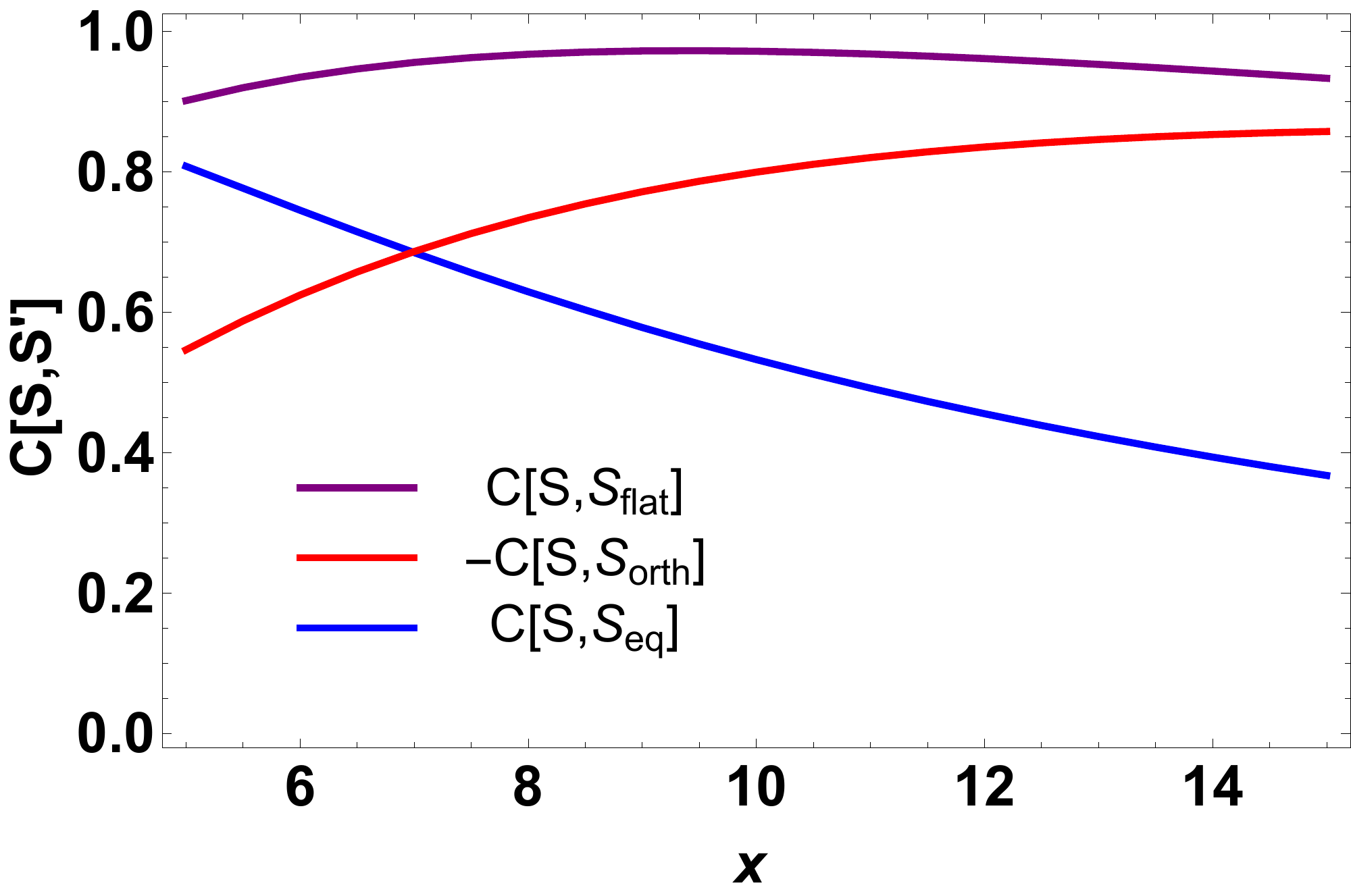}
            \caption{$S_{\dot{\zeta}^3}$}    
            \label{fig:Ssimple-correlation}
        \end{subfigure}
        \hfill
        \begin{subfigure}[b]{0.43\linewidth}  
            \centering 
            \includegraphics[width=\textwidth]{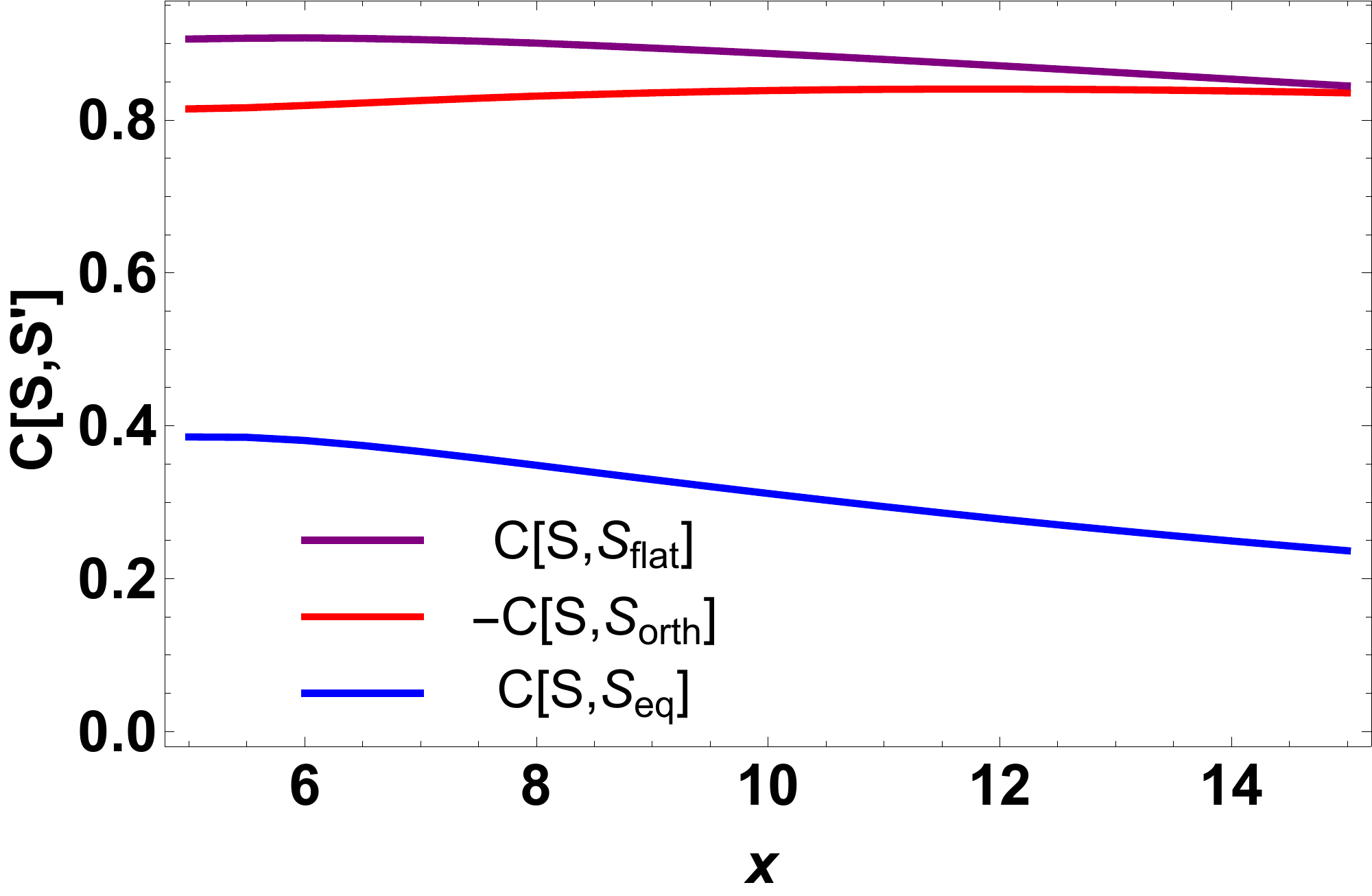}
            \caption{$S_{\dot{\zeta}(\partial\zeta)^2}$}    
            \label{fig:Suni-correlation}
        \end{subfigure}
        \caption{Correlations of $S_{\dot{\zeta}^3}$ with $A>0$ (right), and $S_{\dot{\zeta}(\partial\zeta)^2}$ (left), with the templates in \eqref{templates}, as a function of $\xs$.}
        \label{fig:correlations}
    \end{figure*}

\begin{figure*}
        \centering
        \begin{subfigure}[b]{0.43\linewidth}
            \centering
            \includegraphics[width=\textwidth]{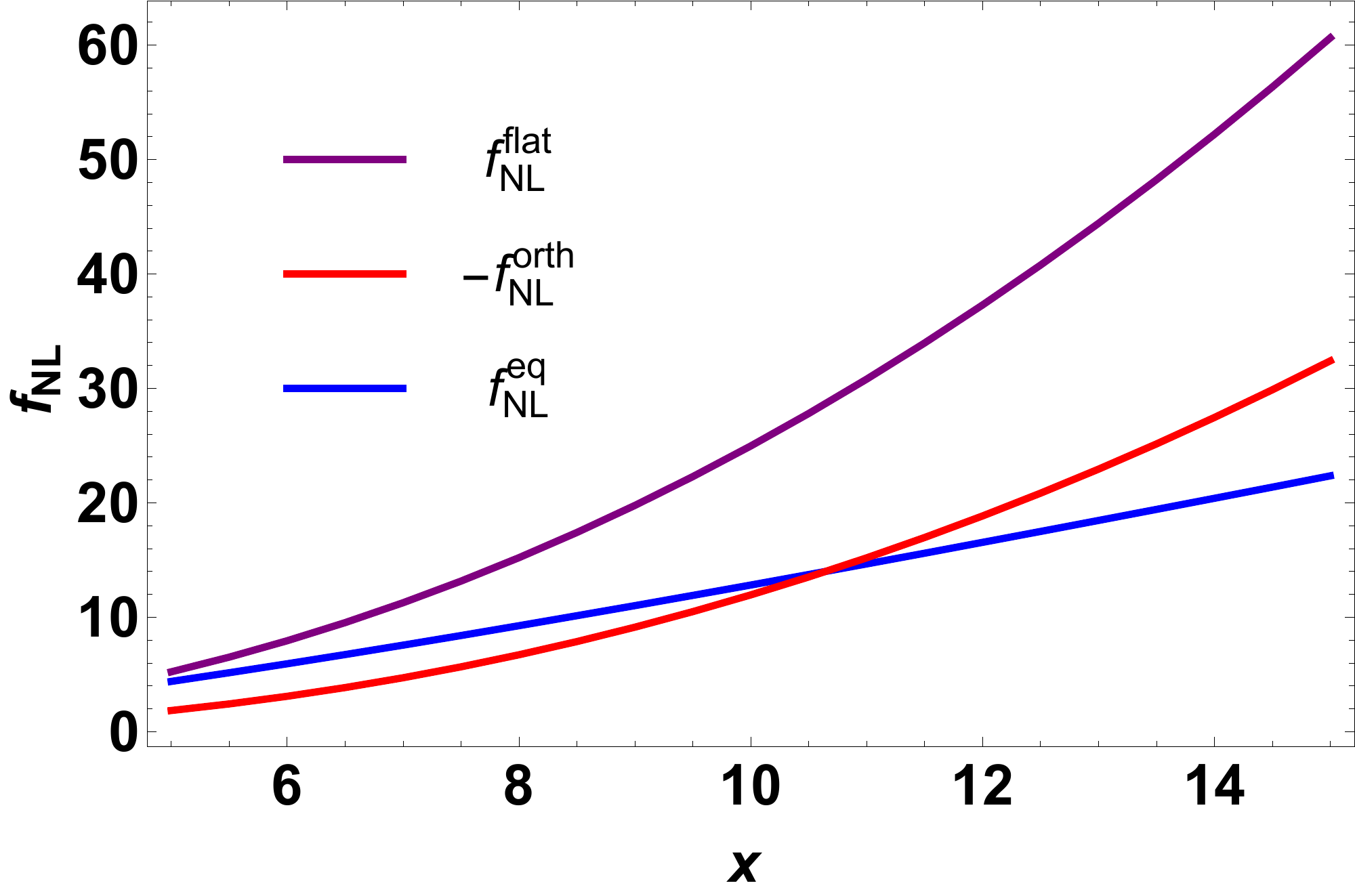}
            \caption{$S_{\dot{\zeta}^3}$}    
            \label{fig:fNLsimple}
        \end{subfigure}
        \hfill
        \begin{subfigure}[b]{0.43\linewidth}  
            \centering 
            \includegraphics[width=\textwidth]{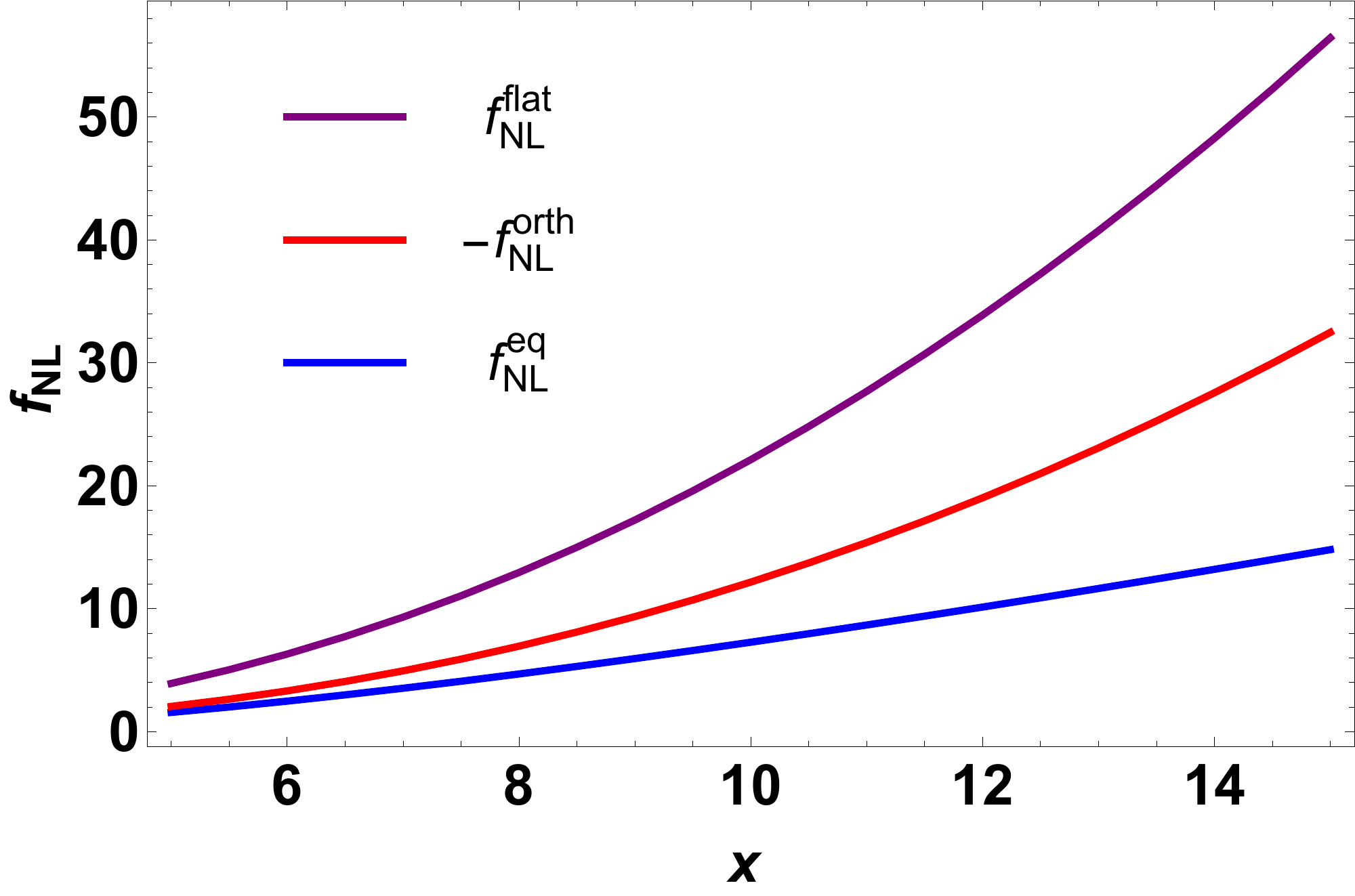}
            \caption{$S_{\dot{\zeta}(\partial\zeta)^2}$}    
            \label{fig:fNLuni}
        \end{subfigure}
        \caption{Amplitudes of the various $f_{NL}$ as a function of $\xs$, for $S_{\dot{\zeta}^3}$ with $A=1$ (right) and $S_{\dot{\zeta}(\partial\zeta)^2}$ (left). The overall common factor $\left(1/|c_s|^2+1 \right)$ is not included.}
        \label{fig:fNL}
    \end{figure*}

We now make a quantitative comparison between the two shapes generated in our set-up with an imaginary speed of sound, and well-known templates used in data analysis, namely the equilateral \cite{Creminelli:2005hu}, orthogonal \cite{Senatore:2009gt}, and flattened (also known as enfolded) \cite{Meerburg:2009ys} shapes:
\begin{equation}
\label{templates}
S^{\eq}=\frac{9}{10} \frac{\kt_1 \kt_2 \kt_3}{k_1 k_2 k_3} \,, \quad S^{\orth}=3\,S^{\eq}-\frac95\,, \quad  S^{\flat}=-S^{\eq}+\frac{9}{10}\,,
\end{equation}
The templates are represented in Fig.~\ref{fig:templates}. 
We make use of the standard inner product $F(S,S')$ \cite{Babich:2004gb,Fergusson:2008ra} (the integral of $S S'$ over the various inequivalent triangle configurations, weighted by $1/(k_1+k_2+k_3)$) to compute the correlation $\mathcal{C}(S,S')$ between a given shape $S$ and a template $S'$, as well as the corresponding amplitude, as
\begin{equation}
\mathcal{C}(S,S')=\frac{F(S,S')}{\sqrt{F(S,S)F(S',S')}}\,, \qquad f_{NL}^{S'}(S)=\frac{F(S,S')}{F(S',S')}\,.
\label{correlations-fNL}
\end{equation}
The results for $S_{\dot{\zeta}^3}$ with $A=1$ and for $S_{\dot{\zeta}(\partial\zeta)^2}$ are represented in Figs.~\ref{fig:correlations} and \ref{fig:fNL} for the correlations and the amplitudes, respectively (factoring out the overall amplitude $\left(1/|c_s|^2+1 \right)$), for $5 \leq \xs \leq 15$. Model-dependent corrections to the universal results that we computed, in $\xs^2 e^{-\xs}$, are not entirely negligible for $5 \lesssim \xs \lesssim 8$, but it is nonetheless interesting to see how our results behave in this regime.

Obviously, the equilateral template, maximum in the equilateral configuration and vanishing in the flattened ones, is a very poor fit to the shapes obtained in our set-up, hence the weak correlations, of order $0.5$ for $S_{\dot{\zeta}^3}$ and $0.2$ for $S_{\dot{\zeta}(\partial\zeta)^2}$. These correlations increase with decreasing $\xs$, as the enhancement of the flattened configurations compared to the equilateral one decreases then. The fact that $S_{\dot{\zeta}(\partial\zeta)^2}$ changes sign between these two types of configurations, whereas $S^\eq$ is always positive, additionally explains its weaker correlation with $S^\eq$ compared to $S_{\dot{\zeta}^3}$. 

The orthogonal template differs from the equilateral one by a constant, so that its amplitude in flattened configurations is $-2$ times the one in the equilateral limit (see Eq.~\eqref{templates} and Fig.~\ref{fig:Seq-orth}). Hence, we expect its correlation with our shapes to be much larger, which is indeed clearly visible in Fig.~\ref{fig:correlations} with correlations of order $-0.8$, this time larger for $S_{\dot{\zeta}(\partial\zeta)^2}$ than for $S_{\dot{\zeta}^3}$, for the same reason that explains the change of sign between equilateral and flattened configurations for the former.

The flattened shape also differs from the equilateral one by a constant, in a way such that the shape is vanishing in the equilateral configuration and maximum in the flattened one (see Eq.~\eqref{templates} and Fig.~\ref{fig:Sflat}). This very well represents the large enhancement of flattened versus equilateral configurations that are typical of our shapes, as one can see in Figs.~\ref{fig:Shapes}-\ref{fig:Sflat}, and which is confirmed quantitatively by the very large correlation of our two shapes with this template, at least of order $0.9$ for all values of $\xs$. This is physically transparent given our computation and explanations in section \ref{computation}: we saw there the important similarities between the non-Gaussianities generated in models with negative $c_s^2$ and the ones induced by non-Bunch--Davies initial states, for which the flattened shape was designed as a simple template \cite{Meerburg:2009ys}.

Finally, one can see in Fig.~\ref{fig:fNL} that the amplitude of the non-Gaussian signal, measured by either of the parameters $f_{NL}^\eq, f_{NL}^\orth, f_{NL}^\flat$, is rather large, independently of the overall common factor $\left(1/|c_s|^2+1 \right)$, and increases with $\xs$. This is due to the polynomial dependence, in $\xs^2$ and $\xs^3$, of the bispectrum near flattened configurations. The largest signal is naturally measured by the flattened template, with $f_{NL}^\flat$ of order $30$ for $\xs=10$ for instance, but even $f_{NL}^\eq={\cal O}(10)$ despite the weak correlation of our shapes with $S^\eq$. Note also that we included the three amplitudes for consistency, but that they are simply related by $f_{NL}^\flat \simeq 0.560 f_{NL}^\eq-1.477 f_{NL}^\orth$.\\

\begin{figure*}[t]
\centering
\includegraphics[width=0.5\textwidth]{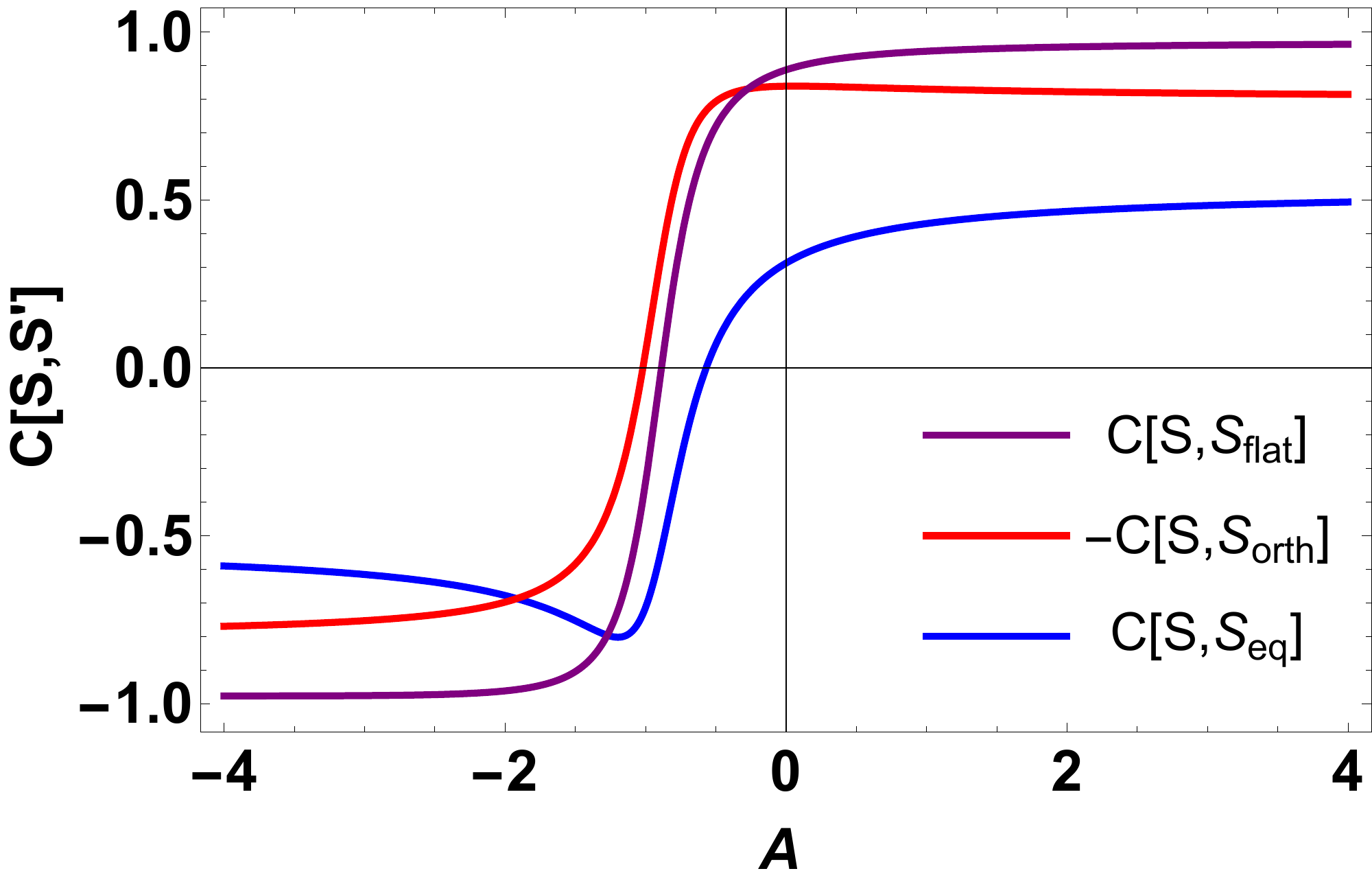}
\caption{Correlation of the total shape $S=S_{\dot{\zeta}^3}+A S_{\dot{\zeta}(\partial\zeta)^2}$ as a function of $A$, for $\xs=10$.}
\label{fig:total10A}
\end{figure*}

\begin{figure*}[t]
\centering
\includegraphics[width=0.5\textwidth]{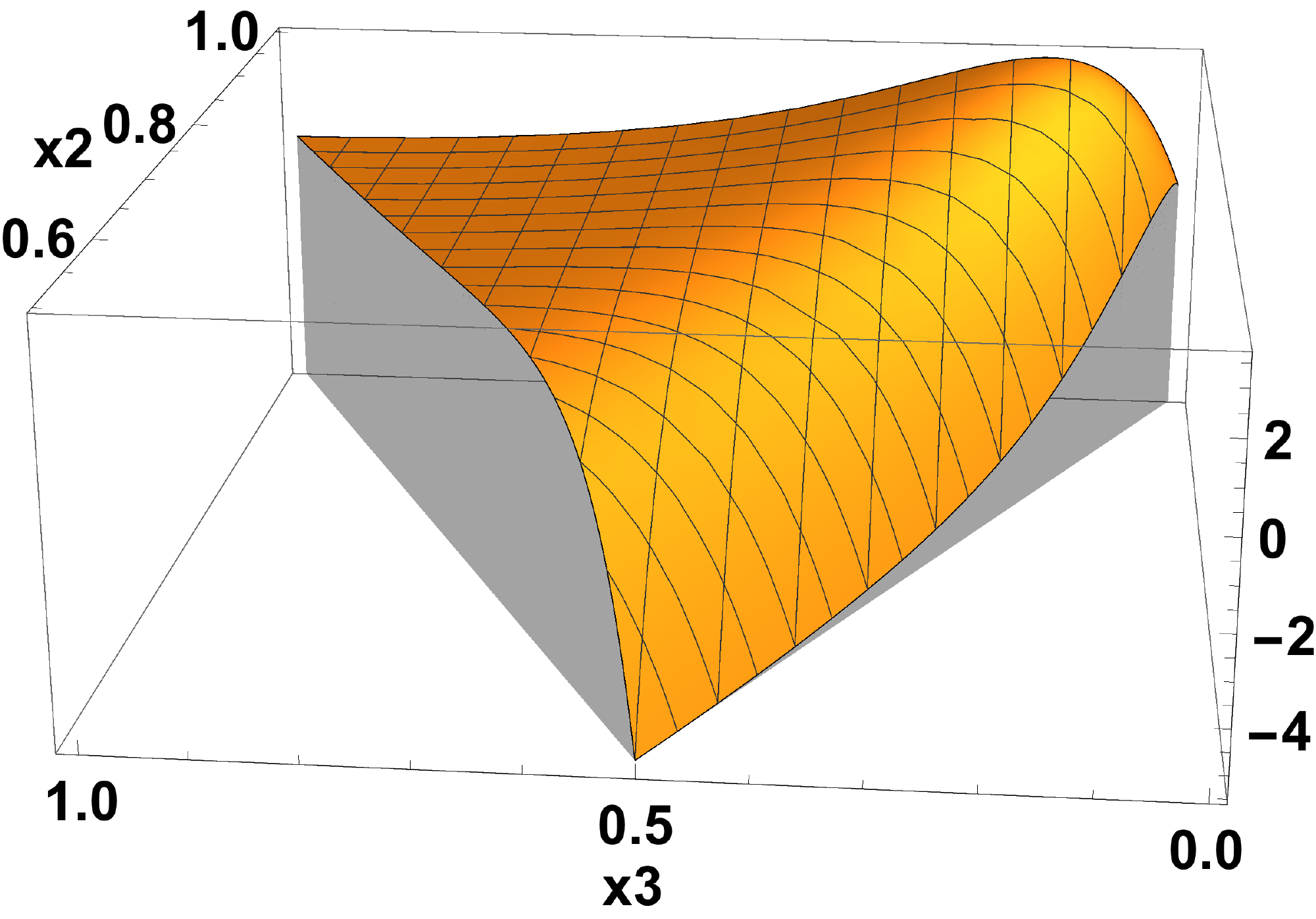}
\caption{Total shape for $\xs=10$ and $A\simeq -0.88$, such that its overlap with the flattened shape is vanishing. It is normalized to $1$ in the equilateral configuration.}
\label{fig:totalmin10}
\end{figure*}

So far we have studied the two shapes independently, but the total bispectrum $S=S_{\dot{\zeta}^3}+A S_{\dot{\zeta}(\partial\zeta)^2}$ is a linear combination of them that depends on the dimensionless parameter $A$, with an amplitude $f_{NL}^X=f_{NL}^X (S_{\dot{\zeta}^3})+A \,f_{NL}^X (S_{\dot{\zeta}(\partial\zeta)^2})$. We show in Fig.~\ref{fig:total10A} how the correlations with the three templates vary as a function of $A$, for the representative value $\xs=10$. As the two individual shapes are strongly correlated with themselves and the flattened template, and have a very comparable amplitude near the most important flattened and squashed configurations (see Eq.~\eqref{S-squashed}), the resulting total shape is either strongly anti-correlated or correlated with the flattened shape, except in a narrow region of parameter space near $A \simeq -1$ (at which the dominant signal in $\xs^3$ is cancelled, see again Eq.~\eqref{S-squashed}). The situation is similar to what happens for positive $c_s^2$ \cite{Senatore:2009gt}, where the two individual shapes there are strongly correlated with the equilateral template, and the total shape is qualitatively different for $3.1 \lesssim A \lesssim 4.2$. This is actually how the orthogonal template was designed, in order not to be blind to this type of signal. In our case, however, one does not need another template, as there always exists a correlation of the total shape (with either $S^\flat$ or $S^\eq$) that is not negligible, and additionally because of the intrinsically large amplitude of the bispectrum. It is nonetheless interesting to represent the total shape for the value of $A$ that generates a vanishing $f_{NL}^\flat$. We do so for $\xs=10$ (with $A \simeq -0.88$ in that case) in Fig.~\ref{fig:totalmin10}, finding similar shapes for different values of $\xs$. Its amplitude is non-negligible near equilateral configurations (which explains why the overlap with $S^\eq$ is non-negligible), but the most important signal is for flattened triangles, with a different sign between the squashed limit and configurations approaching the squeezed limit, near which the shape has a local extremum. In this respect, we note however that in concrete UV realizations of imaginary sound speed scenarios, contributions to the bispectrum coming from times preceding the validity of the EFT might not be entirely negligible near squeezed configurations.

\section{Discussion}
\label{Discussion}

The main purpose of this paper was to work out the consequences of an imaginary speed of sound for the inflationary bispectrum. We considered the simplest effective field theory of fluctuations at lowest order in derivatives, computed the primordial bispectrum and studied its amplitude and shape-dependence. A theory with an imaginary speed of sound cannot be regarded as fundamental but can perfectly make sense as a low energy EFT. In order to make predictions, we thus introduced a physical cut-off momentum scale, parameterized by the dimensionless parameter $\xs$, such that the EFT becomes valid once $\c k/a $ drops below $\xs H$. The parameter $\xs$ measures how deep inside the sound horizon the EFT is trustable (each mode experiences a tachyonic growth during ${\rm ln}(\xs)$ e-folds of expansion between when the scale enters the domain of validity of the EFT and when it exits the sound horizon), and encodes a sensitivity to the ultraviolet completion of the theory.

An imaginary speed of sound induces an instability of the fluctuations, which experience an exponential growth in conformal time, before becoming constant after sound Hubble crossing. However, we showed that the exponentially decreasing mode is essential to a calculation of the non-Gaussianities. Without further input from an UV completion, we worked under the mild assumption that the growing and decaying modes are initially excited with a similar amplitude, which we left unspecified however. Very interestingly, we showed that the dimensionless bispectrum is nonetheless unambiguously determined, at least as soon as $\xs \gtrsim 8$. For this, despite the fact an imaginary speed of sound seems to essentially describe a classical instability, it was important to acknowledge the quantum nature of such a system. It is indeed the commutation relation imposed by the quantization that eventually leads to the determination of the overall amplitude of the bispectrum.

In this respect, it is instructive to compare the two scenarios with positive and negative $c_s^2$. In the former case, the quantization condition determines $|A_k|^2-|B_k|^2$ (see Eq.~\eqref{quantization-standard}), where $A_k$ and $B_k$ are the amplitudes of the positive and negative frequency modes respectively. One is then free to choose the Bunch--Davies vacuum, with $B_k=0$, which unambiguously determines the full mode function, and hence the power spectrum and bispectrum, with equilateral-type non-Gaussianities of amplitude $f_{NL} \sim 1/c_s^2-1$. One can also consider the effect of a small non-Bunch--Davies component in this context, turning on a small non-zero $B_k$. The overall amplitude of the power spectrum is then fixed, but as a result of the interferences between positive and negative frequency modes, the power spectrum comes with superimposed oscillations, whose amplitude is undetermined, but that is tightly constrained observationally \cite{Meerburg:2013dla,Ashoorioon:2013eia,Ade:2015lrj}. In addition to the `standard' part of the bispectrum, the 3-point function also acquires a contribution from the non-Bunch--Davies component, whose amplitude is related to the ones of the power spectrum oscillations, hence intrinsically UV-dependent, and with a shape that is enhanced near flattened configurations, and that features an oscillatory behavior \cite{Chen:2006nt,Holman:2007na,Meerburg:2009ys,Meerburg:2009fi,Agarwal:2012mq}.

The situation with an imaginary speed of sound (negative $c_s^2$) borrows ingredients from the two situations just described. Here, the quantization condition does not determine $|A_k|^2-|B_k|^2$, but rather ${\rm Im}[A_k^* B_k]$ (see Eq.~\eqref{quantization-A-B}), where now $A_k$ and $B_k$ are related to the amplitudes of the exponentially growing and decreasing modes. Hence, one is forced to have a non-zero $B_k$, effectively mimicking a non-Bunch--Davies component. The amplitude of the power spectrum is not determined unambiguously, and the interference between the two types of modes does not result in oscillations, but rather in exponentially suppressed corrections to the leading-order result induced by the growing mode. The amplitude and shape of the dimensionless bispectrum, however, is completely determined, again with $f_{NL} \sim 1/c_s^2-1=-(1/\c^2+1)$ in equilateral configurations, but with a shape that is enhanced by $\xs^3$ near flattened configurations, and lacking features such as the aforementioned oscillations. The bispectrum thus constitutes a more robust probe of imaginary sound speed scenarios than the power spectrum.

We performed a quantitative study of this bispectrum, calculating its correlation with equilateral, orthogonal and flattened templates, as well as the corresponding $f_{NL}$ parameters. The total bispectrum is the sum of two components corresponding to the two cubic vertices of the EFT, and each have a modest correlation with the equilateral shape, but a large correlation with the orthogonal template (of order $0.8)$, and even more so with the flattened one (of order $0.9$). Independently of the overall common factor $\left(1/|c_s|^2+1 \right)$, whose magnitude we left arbitrary, the amplitudes of these shapes are rather large, with $f_{NL}^\flat={\cal O}(30)$, and growing with $\xs$. As the total shape is a linear combination of them, depending on an order one coefficient $A$, we showed that a total shape qualitatively different from its individual flattened-shape components is realized in a narrow region of parameter space near $A \simeq -1$, with only a mild dependence on $\xs$. However, despite its non-standard momentum-dependence, no further template is needed to study it in a first approximation, because of its non-negligible correlation with the equilateral one.

We recently studied concrete UV realizations of imaginary sound speed scenarios in a class of multi-field models that we called sidetracked inflation \cite{Garcia-Saenz:2018ifx}. While it is beyond the scope of this paper to make a detailed comparison, our results here are in qualitative very good agreement with the bispectrum computed there numerically from first principles, with an overall amplitude of the bispectrum set by $1/\c^2+1$, without the exponential enhancement by $e^{2 \xs}$ obtained for the power spectrum, and with a shape that is enhanced in flattened configurations. The main ingredients of the relevant scenarios studied in sidetracked inflation are rather model-independent, and it is useful to make the link between the parameters in such multi-field models and the language that we use here. We refer the reader to reference \cite{Garcia-Saenz:2018ifx} for more details, but the upshot is that imaginary sound speed scenarios arise there as a result of integrating out entropic fluctuations that are heavy and tachyonic, \textit{i.e.}~with the relevant mass parameter such that $m_s^2<0$ and $|m_s|^2 \gg H^2$. As explained there, this type of configuration is compatible with a stable background when the background trajectory deviates strongly from a geodesic, as quantified by the dimensionless parameter $\etaperp$, provided that $m_s^2+4 H^2 \etaperp^2>0$. The transient tachyonic instability experienced by the entropic fluctuations leads to an imaginary speed of sound once they are integrated out, with
\be
\frac{1}{\c^2}+1= \frac{4 H^2 \etaperp^2}{|m_s|^2}\,.
\ee
The description in terms of a single field EFT with a negative $c_s^2$ becomes valid when the physical momenta $k/a$ becomes negligible compared to the mass $|m_s|$ of the field that is integrated out (see appendix \ref{sec:appendix} for details and caveats), thus giving a parametric dependence
\be
\xs \sim  \frac{|m_s|}{H} \c\,,
\ee
where the numerical factor in the right hand side should be somewhat smaller than unity. The two important parameters $\c$ and $\xs$ controlling the EFT are hence determined by $|m_s|/H$ and $\etaperp$ in these UV realizations. In addition, we note that the dominant amplitude of the dimensionless shape function in squashed and flattened configurations scales, both for order one or small $\c$, as $1/\etaperp  (|m_s|/H)^4 < 16 \etaperp^3$. \\

It would be interesting to further investigate the constraints that theoretical consistency puts on imaginary sound speed descriptions. In our motivating example, namely the sidetracked inflationary scenarios of Ref.~\cite{Garcia-Saenz:2018ifx}, we identified interesting attractor homogeneous solutions, and used standard perturbation theory about it to compute two- and three-point correlation functions of cosmological fluctuations. All the salient features observed in this framework are captured by a single-field effective field theory with an imaginary speed of sound, as we showed in \cite{Garcia-Saenz:2018ifx} for the power spectrum, and in this paper for the bispectrum. Another question is to investigate whether the exponential growth of fluctuations in such set-ups, described at the multi-field level, or by an effective single-field theory, can hinder the perturbative approach itself, and which constraints this can put on the parameters of such theories. For instance, requiring that the energy density of the fluctuations be subdominant compared to $H^2 \M^2$, in order to avoid any backreaction, should set an upper bound on $\xs$, in the same way that backreaction constrains excited initial states in models with $c_s^2>0$ (see \textit{e.g.}~\cite{Holman:2007na,Agarwal:2012mq}). It would also be desirable to understand how the discussion in Ref.~\cite{Baumann:2011su} extends to set-ups with imaginary sound speed, as the identification of relevant energy scales might be subtle in scenarios that intrinsically feature instabilities. Without mentioning possibly interesting constraints set by high-order correlation functions, one should at least require $f_{NL} \zeta \ll 1$ for the perturbative description to be valid. Using the power spectrum \eqref{As} as an estimate for the amplitude of $\zeta \sim A_s^{1/2} = \alpha e^{\xs}/(\sqrt{2} \pi)$, and omitting numerical factors, one finds $\left(\frac{1}{\c^2}+1 \right) x^3 \alpha e^{\xs} \ll 1$\,.
Despite Eq.~\eqref{quantization-apha}, the amplitude of $\alpha$ itself is not specified in terms of the parameters of our EFT, as it depends on the specific UV completion of it, but one can envisage to add other operators to extend its regime of validity, for instance along the lines of Refs.~\cite{Baumann:2011su,Gwyn:2012mw,Gwyn:2014doa}. Finally, as it is clear from Eq.~\eqref{S-pi}, having $c_s^2<0$ with $\epsilon>0$ and having  $c_s^2>0$ with $\epsilon<0$ equally implies a negative kinetic energy of $\pi$. In spite of the differences between the two set-ups, it would hence be interesting to compare imaginary sound speed models with systems that violate the null energy condition. It would also be intriguing to understand whether constraints derived in other contexts on low-energy effective ghosts, for instance related to their decay into gravitons \cite{Carroll:2003st,Cline:2003gs}, can be used to further constrain the framework studied here. We leave these various questions, as well as further studies of concrete realizations of imaginary sound speed scenarios, for future works.

\begin{acknowledgments}

We are grateful to Patrick Peter, Lucas Pinol, John Ronayne and Krzysztof Turzy\'nski for useful discussions. S.GS is supported by the European Research Council under the European Community's Seventh Framework Programme (FP7/2007-2013 Grant Agreement no.\ 307934, NIRG project). S.RP is supported by the European Research Council (ERC) under the European Union's Horizon 2020 research and innovation programme (grant agreement No 758792, project GEODESI). 

\end{acknowledgments}

\appendix

\section{Effective single-field action from a two-field model} \label{sec:appendix}

In this appendix we give the explicit derivation of the second-order effective action for the adiabatic fluctuations obtained from integrating out the entropic mode in a two-field model of inflation. Our goal is to show how the effective single-field theory \eqref{eq:effective goldstone action} that we focus on in this work (but restricted here to second order for simplicity) can arise from a concrete UV completion, and in particular that the speed of sound can be imaginary even if the full theory has no fundamental pathologies. We present this for the sake of clarity and completeness, as analogous derivations have already been given in several previous works (see {\it e.g.}~\cite{Tolley:2009fg,Cremonini:2010ua,Achucarro:2010da}). The theory we consider is a generic non-linear sigma model of the form
\beq \label{eq:app-full action}
S[\phi^1,\phi^2]=\int d^4x\sqrt{-g}\bigg[-\frac{1}{2}\,G_{IJ}(\phi)\nabla^{\mu}\phi^I\nabla_{\mu}\phi^J-V(\phi)\bigg]\,,
\eeq
where the potential and the field-space metric $G_{IJ}$ are arbitrary. We consider a background where the metric is spatially flat and of the FLRW type, and where the fields are homogeneous. To study perturbations above such a background, it is convenient for our purposes to use the adiabatic/entropic fluctuations, with Fourier-space variables $v_\sigma=a \,Q_\sigma$ and $v_s=a \,Q_s$, where $Q_\sigma$ and $Q_s$ are the gauge invariant variables that reduce in the spatially flat gauge to the field fluctuations along and orthogonal to the background trajectory, respectively. The second-order action for these gauge-invariant fluctuations reads (see \cite{Langlois:2008mn,Langlois:2008qf} for computations that encompass this case but hold for a more general class of theories than the one in \eqref{eq:app-full action}):
\beq\bal
S^{(2)}[\zeta=z v_\sigma,v_s]&=\frac{1}{2}\int d\tau d^3k\bigg[  z^2 \left( \zeta^{\prime2}-k^2 \zeta^2  \right)  +v_s^{\prime2}-k^2v_s^2\\
&\quad+\left(\frac{a''}{a^3}-m_s^2\right)a^2v_s^2 -2\xi z v_s \zeta^{\prime} \bigg]\,.
\label{S2-2fields}
\eal\eeq
Here primes denote derivatives with respect to conformal time $\tau$ and one uses the curvature perturbation $\zeta = z v_\sigma$. We have also introduced the definitions
\beq
z\equiv a\sqrt{2\epsilon}\,,\qquad \xi\equiv 2aH\eta_{\perp}\,,
\eeq
where $\epsilon=-\dot{H}/H^2$ is the slow-roll parameter, $\eta_\perp=-V_{,s}/(H^2M_{\rm Pl}\sqrt{2\epsilon})$ is the so-called bending parameter, and $V_{,s}$ is the derivative of the potential projected along the entropic direction. The parameter $\eta_\perp$ quantifies the deviation of the background trajectory from a field-space geodesic, and is responsible for the coupling between adiabatic and entropic modes at the linear level. Lastly, the entropic mass parameter $m_s^2$ is defined as
\beq
m_s^2\equiv V_{;ss}+\epsilon R_{\rm fs}H^2M_{\rm Pl}^2-\eta_\perp^2H^2\,,
\eeq
with $V_{;ss}$ the second (field-space covariant) entropic derivative of the potential and $R_{\rm fs}$ the Ricci scalar of the internal field space.

We now assume that all background quantities evolve much less rapidly than the scale factor, that the entropic mode is heavy, in the sense that $|m_s^2|\gg H^2$, and restrict to the regime where $k^2/a^2\ll |m_s^2|$. In this regime, we can neglect the kinetic term of $v_s$ (something we will verify and delineate the self-consistency of which afterwards), upon which integrating out $v_s$ by solving for it from its equation of motion readily gives
\beq
v_s=-\frac{z \xi/a^2}{m_s^2} \zeta^{\prime}\,,
\label{vs-zetaprime}
\eeq
where the assumed near de Sitter behavior allows us to neglect $a''/a^3 \simeq 2 H^2$ compared to $m_s^2$ in the denominator.
Substituting back \eqref{vs-zetaprime} in $S^{(2)}$ of \eqref{S2-2fields} then yields an effective second-order action for the curvature perturbation
\beq \label{eq:app-eff action 2}
S^{(2)}_{\rm eff}=\frac12 \int d\tau d^3k\,z^2 \bigg[\frac{\zeta^{\prime2}}{c_s^2}-k^2\,\zeta^2\bigg]\,,
\eeq
where the speed of sound (squared) $c_s^2$ is given by
\beq
\frac{1}{c_{s}^2}\equiv 1+\frac{4H^2\eta_\perp^2}{m_s^2}\,.
\label{cs2}
\eeq
As explained in section \ref{sec:set-up}, at leading order in perturbations and in the slow-roll approximation we have the simple relation $\zeta=-H\pi$ between $\zeta$ and the Goldstone mode $\pi$. The second-order part of the effective Goldstone action \eqref{eq:effective goldstone action} then immediately follows from \eqref{eq:app-eff action 2}, as we wished to show.\\

As announced, one can check the consistency of neglecting the kinetic term of $v_s$ in \eqref{S2-2fields} in the procedure of integrating it out. For this, one can use the $v_s$ deduced from \eqref{vs-zetaprime} and the solution for $\zeta$ resulting from the effective action \eqref{eq:app-eff action 2}, and check that $v_s^{\prime2}$ is negligible compared to the dominant term $a^2 m_s^2 v_s^2$. The starting point for this is the general solution \eqref{general-solution}, which is applicable for $c_s^2$ positive or negative, and from which two physically distinct regimes should be discussed:\\

\textbullet\, Inside the `sound horizon', when $| k c_s \tau| \gtrsim 1$: there one finds $\zeta^\prime \sim k c_s \, \zeta$, hence $v_s^\prime \sim k c_s v_s$, and $v_s^{\prime2}/(a^2 m_s^2 v_s^2) \sim \frac{k^2 c_s^2}{a^2 m_s^2}$, which is the expected ratio between the kinetic and the potential energies inside the sound horizon. 

\textbullet\, Outside the `sound horizon', when $| k c_s \tau| \lesssim 1$: there one finds $\zeta^\prime \sim k^2 c_s^2 \tau \, \zeta$, hence $v_s^\prime \sim a H v_s$, and $v_s^{\prime2}/(a^2 m_s^2 v_s^2) \sim \frac{H^2}{m_s^2}$, which is much smaller than unity for the large entropic mass $|m_s^2| \gg H^2$ that we consider.\\

The self-consistency of the procedure thus requires that 
\beq
\frac{k^2 c_s^2}{a^2 m_s^2} \ll 1\,,
\label{validity}
\eeq
in addition to the physical momentum (squared) $k^2/a^2$ dropping below $m_s^2$. In the standard situation in which $m_s^2 >0$, one has from \eqref{cs2} that $0 < c_s^2 <1$, \textit{i.e.}\ a positive and reduced speed of sound, in which case the bound \eqref{validity} is automatically satisfied as soon as $k^2/a^2$ becomes negligible compared to $m_s^2$. 

In our case of interest in this paper, with negative $m_s^2$, $c_s^2$ is negative, and can \textit{a priori} be larger than one. If it is smaller than unity, then the bound \eqref{validity} is again automatically satisfied as soon as $k^2/a^2$ becomes negligible compared to $|m_s^2|$. The regime of validity of the EFT is however delayed if $|c_s^2| \gtrsim 1$, although it is far from clear whether this possibility can indeed arise in a self-consistent manner in a UV complete theory and may presumably be ruled out by positivity constraints \cite{Adams:2006sv} (notice from \eqref{cs2} that this arises only when the two mass scales $|m_s^2|$ and $4 H^2 \eta_\perp^2$ coincide to a high precision). Put the other way around, the EFT becomes applicable as soon as $k^2/a^2$ becomes negligible compared to $|m_s^2|$ only in models with $|c_s^2| \lesssim 1$. In this respect, note that in the type of two-field models identified in Ref.~\cite{Garcia-Saenz:2018ifx} whose single-field effective theory displays an imaginary sound speed, there is a hierarchy $|m_s^2| \ll H^2 \eta_\perp^2 $, so that $|c_s^2| \ll 1$.

From the point of view of the low-energy single-field EFT, we have seen in the main text that a central role is played by the dimensionless parameter $x$, which is the value below which $k |c_s|/(aH)$ should drop for the EFT to be applicable. From \eqref{validity}, we thus find that, when this EFT emerges from integrating out entropic fluctuations along the lines above, theoretical self-consistency imposes the interesting upper bound
\beq
x^2 \ll |m_s^2|/H^2\,.
\eeq


\bibliographystyle{apsrev4-1}
\bibliography{Biblio-2018}

\end{document}